\documentclass[12pt,preprint]{aastex}

\usepackage{graphicx}
\usepackage{color}
\usepackage{ textcomp }

\slugcomment{Revised 2017 March 26}

\shorttitle{P/2013 R3}
\shortauthors{Jewitt et al.}

\begin{document}

\title{Anatomy of an Asteroid Break-Up: The Case of P/2013 R3}
\author{David Jewitt$^{1,2}$, Jessica Agarwal$^3$, Jing Li$^{1}$, Harold Weaver$^4$,  Max Mutchler$^5$ and Stephen Larson$^6$
}
\affil{$^1$Department of Earth, Planetary and Space Sciences,
UCLA, 
595 Charles Young Drive East, 
Los Angeles, CA 90095-1567\\
$^2$Dept.~of Physics and Astronomy,
University of California at Los Angeles, \\
430 Portola Plaza, Box 951547,
Los Angeles, CA 90095-1547\\
$^3$ Max Planck Institute for Solar System Research, Justus-von-Liebig-Weg 3, 37077 G\"ottingen, Germany\\
$^4$ The Johns Hopkins University Applied Physics Laboratory, 11100 Johns Hopkins Road, Laurel, Maryland 20723  \\
$^5$ Space Telescope Science Institute, 3700 San Martin Drive, Baltimore, MD 21218 \\
$^6$ Lunar and Planetary Laboratory, University of Arizona, 1629 E. University Blvd.
Tucson AZ 85721-0092 \\
}

\email{jewitt@ucla.edu}

\begin{abstract}
We present an analysis of new and published data on P/2013 R3, the first asteroid detected while disintegrating. Thirteen  discrete components are measured in the interval between UT 2013 October 01 and 2014 February 13.  We determine a mean, pair-wise velocity dispersion amongst these components of $\Delta v = 0.33\pm0.03$ m s$^{-1}$ and find that their separation times are staggered over  an interval of $\sim$5 months.  Dust enveloping the system has, in the first observations, a  cross-section $\sim$30 km$^2$ but fades monotonically at a rate consistent with the action of radiation pressure sweeping.  The individual components exhibit comet-like morphologies and also fade except where secondary fragmentation is accompanied by the release of additional dust.  We find only upper limits to the radii of any embedded solid nuclei, typically $\sim$100 to 200 m (geometric albedo 0.05 assumed).  Combined, the components of P/2013 R3 would form a single spherical body with radius $\lesssim$400 m, which is our best estimate of the size of the precursor object.  The observations are consistent with rotational disruption of a weak (cohesive strength $\sim$50 to 100 N m$^{-2}$) parent body,  $\sim$400 m in radius. Estimated radiation (YORP) spin-up times of this  parent are $\lesssim$1 Myr, shorter than the collisional lifetime.  If present,  water ice sublimating at as little as 10$^{-3}$ kg s$^{-1}$ could generate a torque on the parent body rivaling the YORP torque.    Under conservative assumptions about the frequency of similar  disruptions, the inferred asteroid debris production rate is $\gtrsim$10$^3$ kg s$^{-1}$, which is at least 4\% of the rate needed to maintain the Zodiacal Cloud.
\end{abstract}


\section{INTRODUCTION}

Remarkable main-belt object P/2013 R3 (hereafter ``R3'') orbits in the asteroid belt but, unlike other asteroids, consists of multiple discrete components enshrouded in a dust envelope (Jewitt et al.~2014).   The orbital semi-major axis, eccentricity and inclination of R3 are 3.033 AU, 0.273 and 0.90\degr, respectively, giving a Tisserand parameter measured with respect to Jupiter of $T_J$ = 3.183. This substantially exceeds the  dividing line between comets (which have $T_J <$ 3)  and asteroids ($T_J > 3$, e.g.~Kosai 1992). The combination of an asteroid-like orbit with a dusty, comet-like appearance qualifies R3 as a member of the active asteroid population  (Jewitt et al.~2015a), in which the causes of  mass loss are many and varied.  About one quarter of the $\sim$20 known examples are observed to be repetitively active when near perihelion, just as are the Kuiper belt and Oort cloud comets.  Repetitive activity in the active asteroids, as in the comets,  is most simply explained as reflecting the sublimation of near surface ice,  although there is no efficient dynamical path from either comet reservoir to the asteroid belt (c.f.~Hsieh and Jewitt 2006, Jewitt 2012, Jewitt et al.~2015a).  Other examples result from the ejection of debris following asteroid-asteroid impact (e.g.~Ishiguro et al.~2011) while others, including R3 and 311P/2013 P5, show evidence suggestive of rotational break-up (Jewitt et al.~2013a, 2014, Drahus et al.~2015, Sheppard and Trujillo 2015, Hui et al.~2017).  In many other active asteroids, the cause of the activity remains indeterminate.  The distinctive appearance of R3, however, remains unique in showing that it is actively disintegrating (Jewitt et al.~2014).  As such, it may be a main-belt analog of the split comets occasionally observed in the classical comet population (e.g.~Boehnhardt 2004).

Our initial paper on R3 (Jewitt et al.~2014) addressed a sub-set of the now-available data.  Here, we present new observations, together with a re-analysis of data discussed in the earlier work.  Our results  substantially confirm and extend those of the earlier paper. In the few instances where differences exist, the present results should be taken to supercede those previously published.  Our science objective is to provide the definitive characterization of R3 and to understand the nature and cause of its disintegration.  

\section{FACILITIES} 

\begin{figure}
\epsscale{0.95}
\begin{center}
\plotone{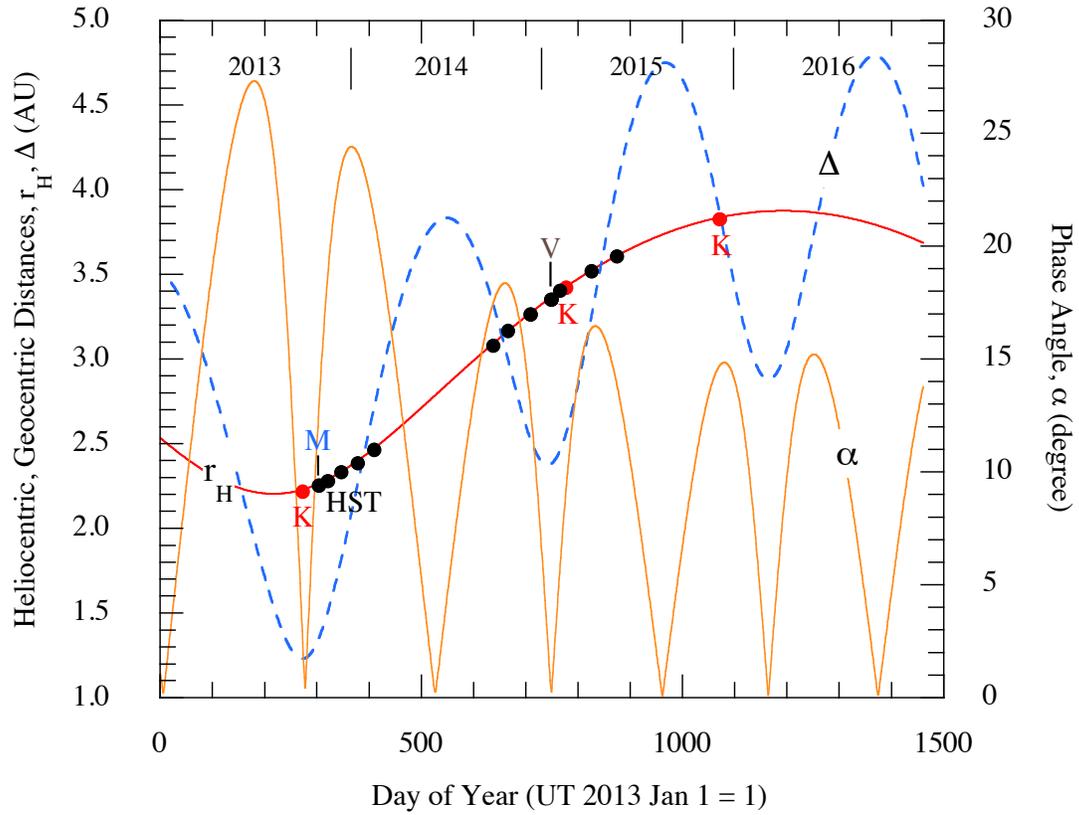}
\caption{Observing geometry, showing heliocentric distance, $r_H$ (red line), geocentric distance, $\Delta$ (dashed blue line) and phase angle, $\alpha$ (orange line).  Filled circles show the dates of the HST, Keck (K), Magellan (M) and VLT (V) observations from Table (\ref{geometry}).   \label{RDa}
} 
\end{center} 
\end{figure}

Observations were obtained using the 2.4 m Hubble Space Telescope, the 10 m Keck telescope, the 6.5 m Magellan telescope and the 8 m Very Large Telescope.   Data from these facilities are highly complementary, by offering a range of angular resolutions and sensitivities on a wide range of observing dates.  The geometrical circumstances are given for each date of observation in Table (\ref{geometry}), while  Figure (\ref{RDa}) shows the time dependence of the heliocentric and geocentric distances and of the phase angle.  In the Table and the Figure, time is given for convenience as Day of Year, such that UT 2013 January 01 = 1.

\subsection{Hubble Space Telescope}

Observations with the Hubble Space Telescope (HST) were taken under GO programs 13612 and 13865.  We used the UVIS channel of the WFC3 camera\footnote{\url{$http://www.stsci.edu/hst/wfc3/documents/handbooks/currentIHB/wfc3_cover.html$}} whose 0.04\arcsec~wide pixels each correspond to about 41 km at the minimum distance of R3 ($\Delta$ = 1.338 AU on UT 2013 October 29), giving a Nyquist-sampled spatial resolution of $\sim$82 km.  The field of view is approximately 162\arcsec$\times$162\arcsec. All observations were taken using the very broad F350LP filter (4758\AA~full width at half maximum, ``FWHM'') which has an effective wavelength of 6230\AA~on a solar-type (G2V) source.  From each orbit in GO 13612 we obtained five exposures of 348 s duration and one of 233 s while in GO 13865 we obtained five exposures each 420 s in duration, per orbit.
 
 \subsection{Keck Telescope}
We used the 10 m diameter Keck 1 telescope atop Mauna Kea equipped with the LRIS camera (Oke et al.~1995).  LRIS provides simultaneous imaging with red and blue sensitive detectors, separated by a dichroic filter.  The image scale on both is 0.135\arcsec~per pixel, giving an unvignetted field approximately 6\arcmin~across.  We employed  broadband B, V and R filters having central wavelengths (and FWHM) 4370\AA~(878\AA), 5473\AA~(948\AA) and 6417\AA~(1185\AA), respectively.  The telescope was tracked at non-sidereal rates in order to follow the expected motion of R3, and the pointing position was dithered to provide protection from chip defects.   The data were internally calibrated using bias frames and flat-field frames, the latter formed from a composite of images of an illuminated patch inside the telescope dome.   Photometric calibration was obtained from images of nearby standard stars (Landolt 1992) and is accurate to $\pm$0.02 magnitudes.

Observations on UT October 1 and 2 were taken to establish the initial  appearance of R3.   In particular, on October 2 a sequence of 22 images in both B and R filters (total integration times of 6600 and 5720 s, respectively) was used to measure the dust distribution around R3, down to low surface brightness levels.  Later LRIS observations on UT 2015 February 17 and December 08 were taken to try to identify the end state of the disintegration of R3.

\subsection{Magellan Telescope}
 The Magellan I (``Baade'') 6.5 m telescope was used to image R3 on UT 2013 October 28 and 29.  These observations were generously taken on our behalf by Scott Sheppard of the Carnegie Institute for Science.  The observations employed the IMACS imaging camera (Dressler et al.~2011) in short focus mode, which gives a 27.2\arcmin~square field of view and an image scale of 0.2\arcsec~per pixel.  We obtained four integrations of 90 s each on October 28 and a further three on UT October 29, all through a Sloan r filter (central wavelength $\sim$6200\AA, FWHM 1390\AA).   The seeing was 0.7\arcsec~FWHM.

\subsection{Very Large Telescope}
On 2015 January 18, we observed the predicted position of R3 with the FOcal Reducer/low dispersion Spectrograph 2 (FORS2) instrument (Appenzeller et al.~1998) mounted on the 8.2m Very Large Telescope (VLT) UT1 (Antu) telescope of the European Southern Observatory (ESO) on Cerro Paranal in Chile. We obtained 30 images of 530\,s exposure time each, in the R\_SPECIAL+76 filter band, which has an effective wavelength of 6550\AA~and  a bandwidth of 1650\AA. The images have a pixel scale of 0.25\arcsec\ (2x2 binning) and a field of view of 6.8\arcmin\ $\times$ 6.8\arcmin. The seeing varied between 0.8\arcsec\ and 1.5\arcsec.

\section{MORPHOLOGY}

\begin{figure}
\epsscale{0.95}
\begin{center}
\includegraphics[width=0.95\textwidth, angle =0 ]{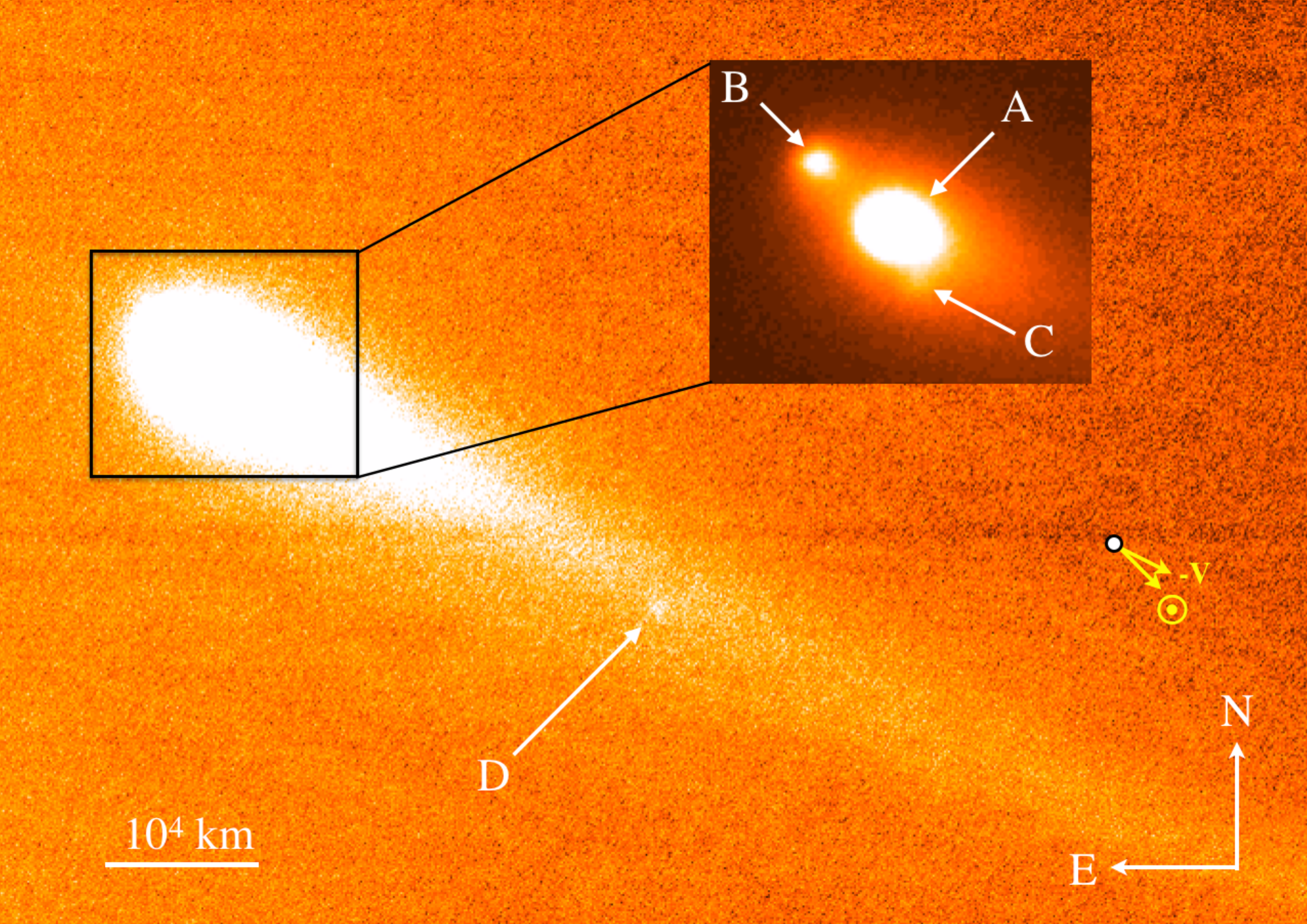}

\caption{Composite of 22 R-band Keck images taken 2013 October 02 and having total exposure time 5720 s.    The major components are labelled in the inset, and the down-tail component D is also marked.  Image has North to the top, East to the left. The position angles of the antisolar vector (marked $\odot$) and the negative heliocentric velocity vector (marked $-V$) are shown in yellow. \label{october02}
} 
\end{center} 
\end{figure}

\begin{figure}
\epsscale{0.95}
\begin{center}
\includegraphics[width=0.95\textwidth, angle =0 ]{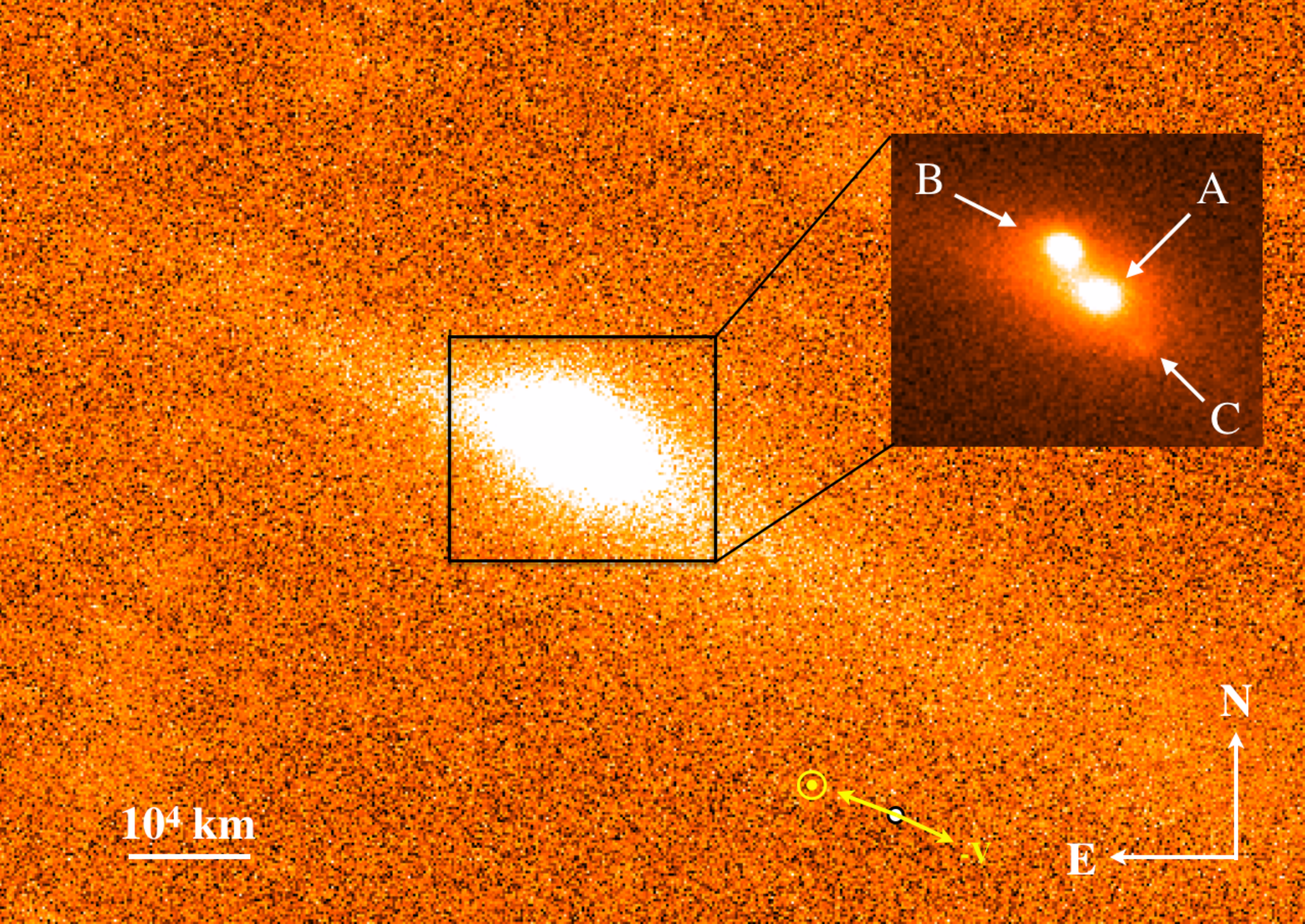}

\caption{Composite of 4 R-band Magellan images taken 2013 October 28 and having total exposure time 360 s.    The major components are labelled; the down-tail component D is too faint to be seen in this image.  Image has North to the top, East to the left. The position angles of the antisolar vector (marked $\odot$) and the negative heliocentric velocity vector (marked $-V$) are shown in yellow. \label{october28}
} 
\end{center} 
\end{figure}

\begin{figure}
\epsscale{0.95}
\begin{center}
\plotone{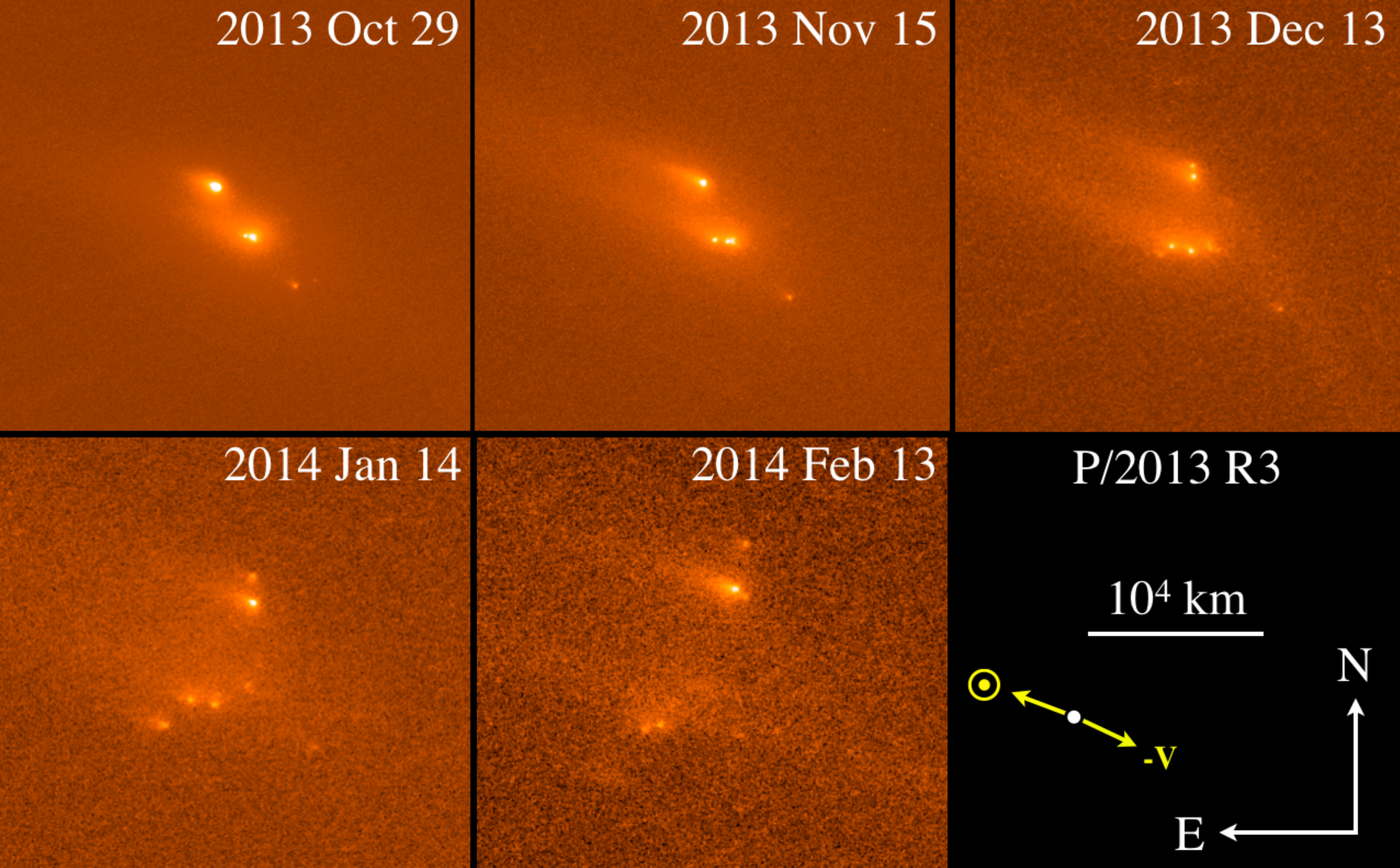}
\caption{Five epochs of HST imaging of R3 from 2013 and 2014 (c.f.~Table \ref{geometry}).  Each panel has North to the top, East to the left.  The images have been scaled according to geocentric distance to give a fixed linear scale, as shown.  The position angles of the antisolar vector (marked $\odot$) and the negative heliocentric velocity vector (marked $-V$) are shown in yellow; these angles change negligibly over the range of dates shown.    \label{whole}
} 
\end{center} 
\end{figure}

\begin{figure}
\epsscale{0.85}
\begin{center}
\includegraphics[width=0.8\textwidth, angle =0 ]{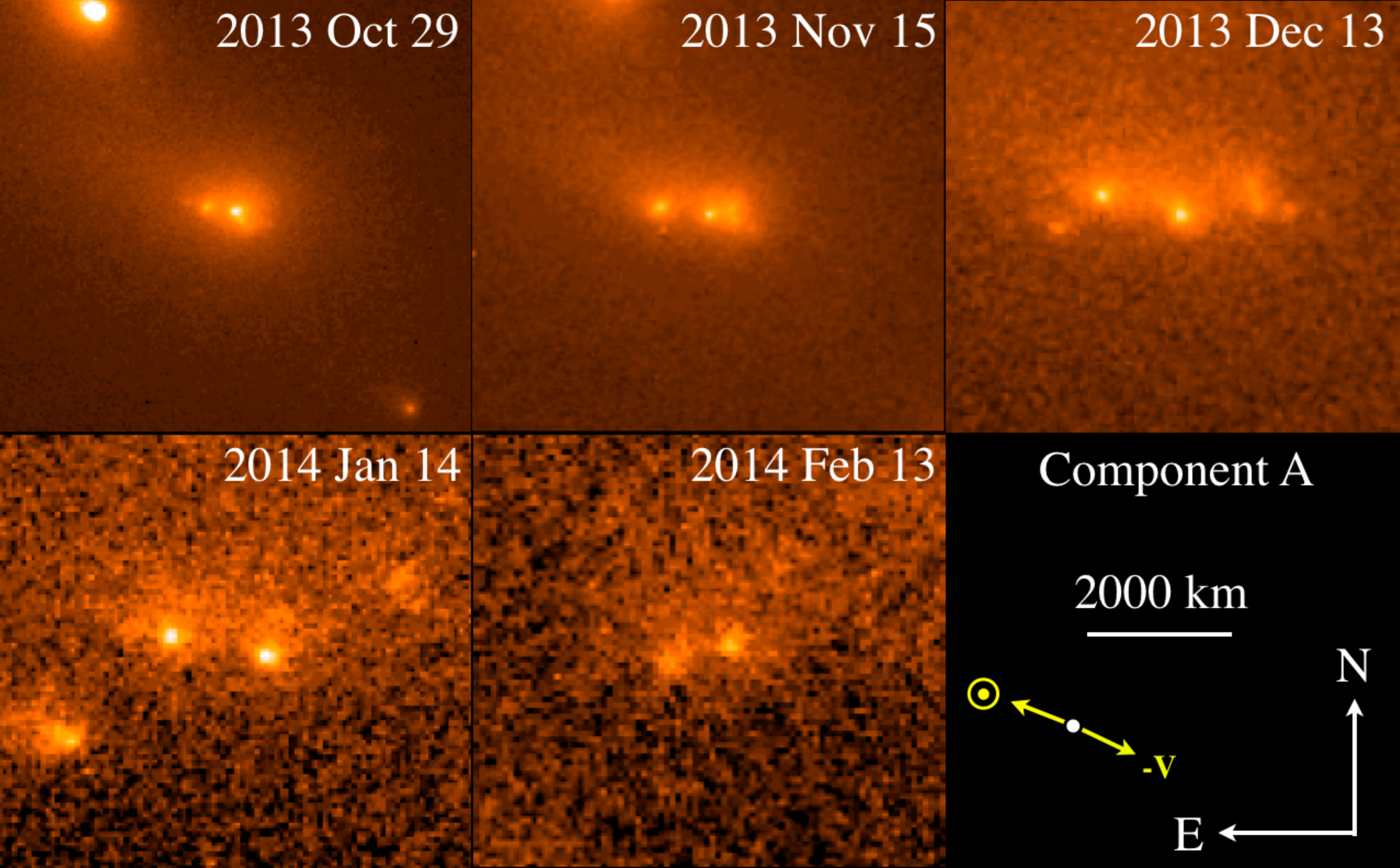}
\includegraphics[width=0.8\textwidth, angle =0 ]{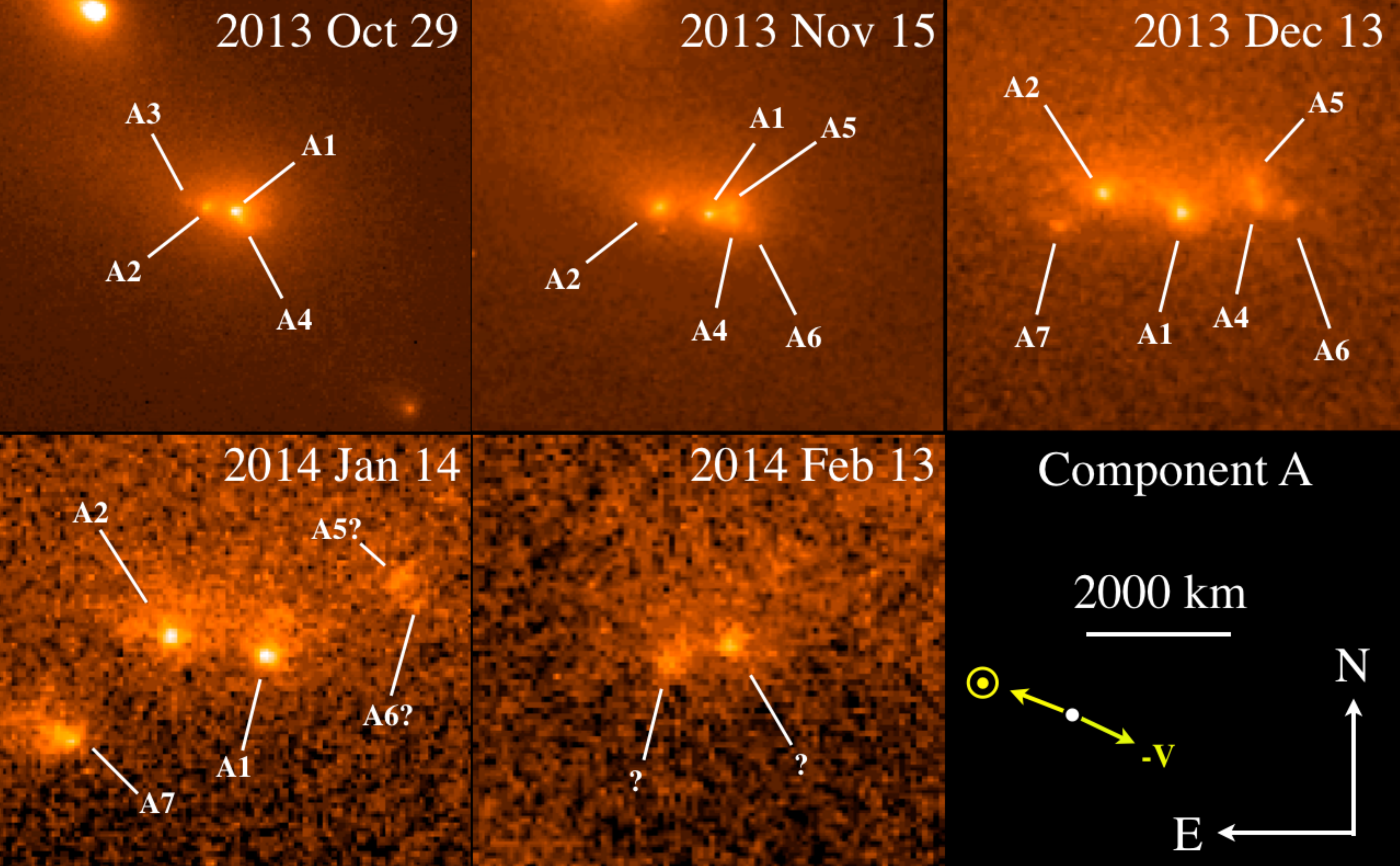}
\caption{Same as Figure (\ref{whole}) but zoomed on component A.   \label{component_A}
} 
\end{center} 
\end{figure}

\begin{figure}
\epsscale{0.85}
\begin{center}
\includegraphics[width=0.8\textwidth, angle =0 ]{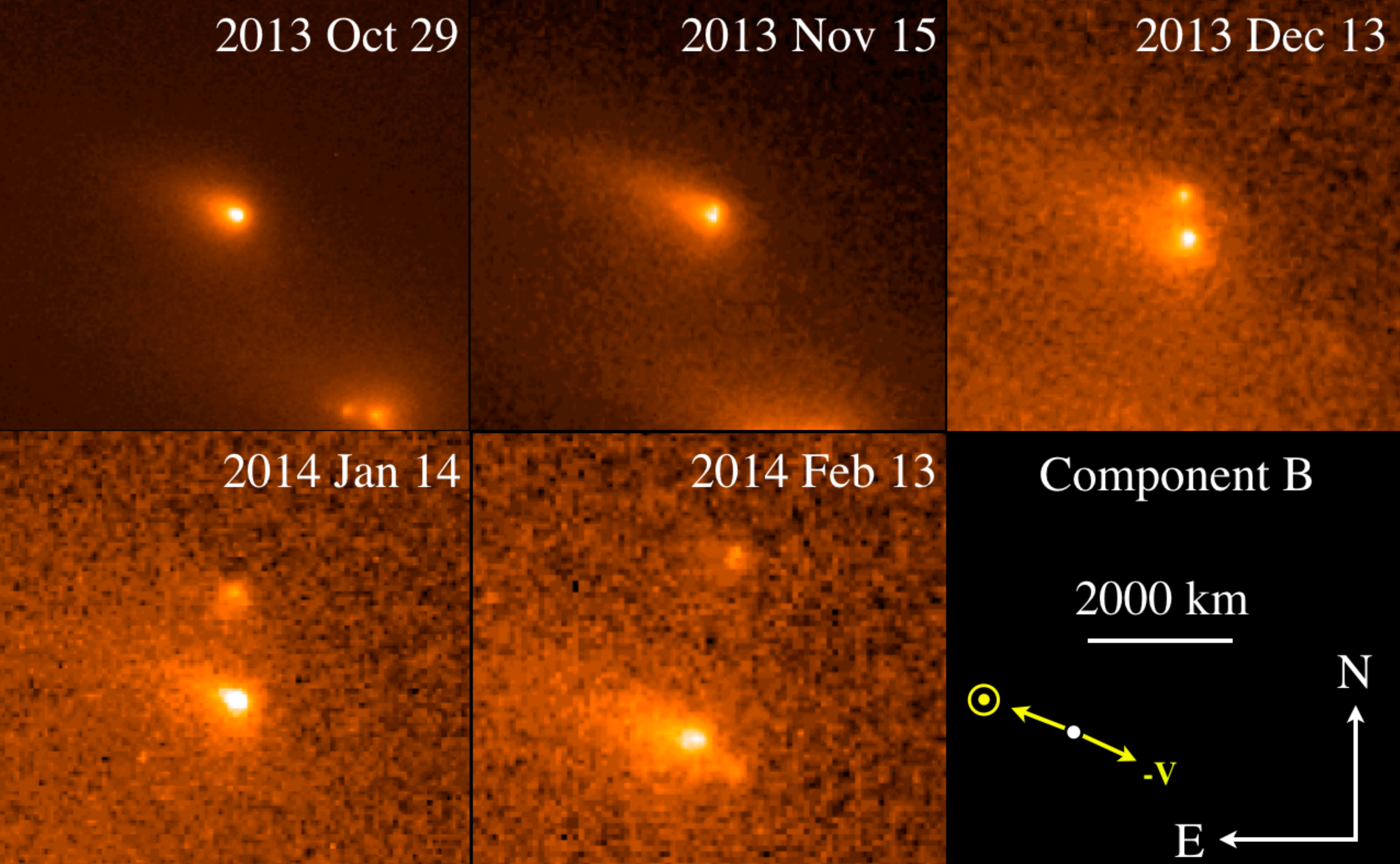}
\includegraphics[width=0.8\textwidth, angle =0 ]{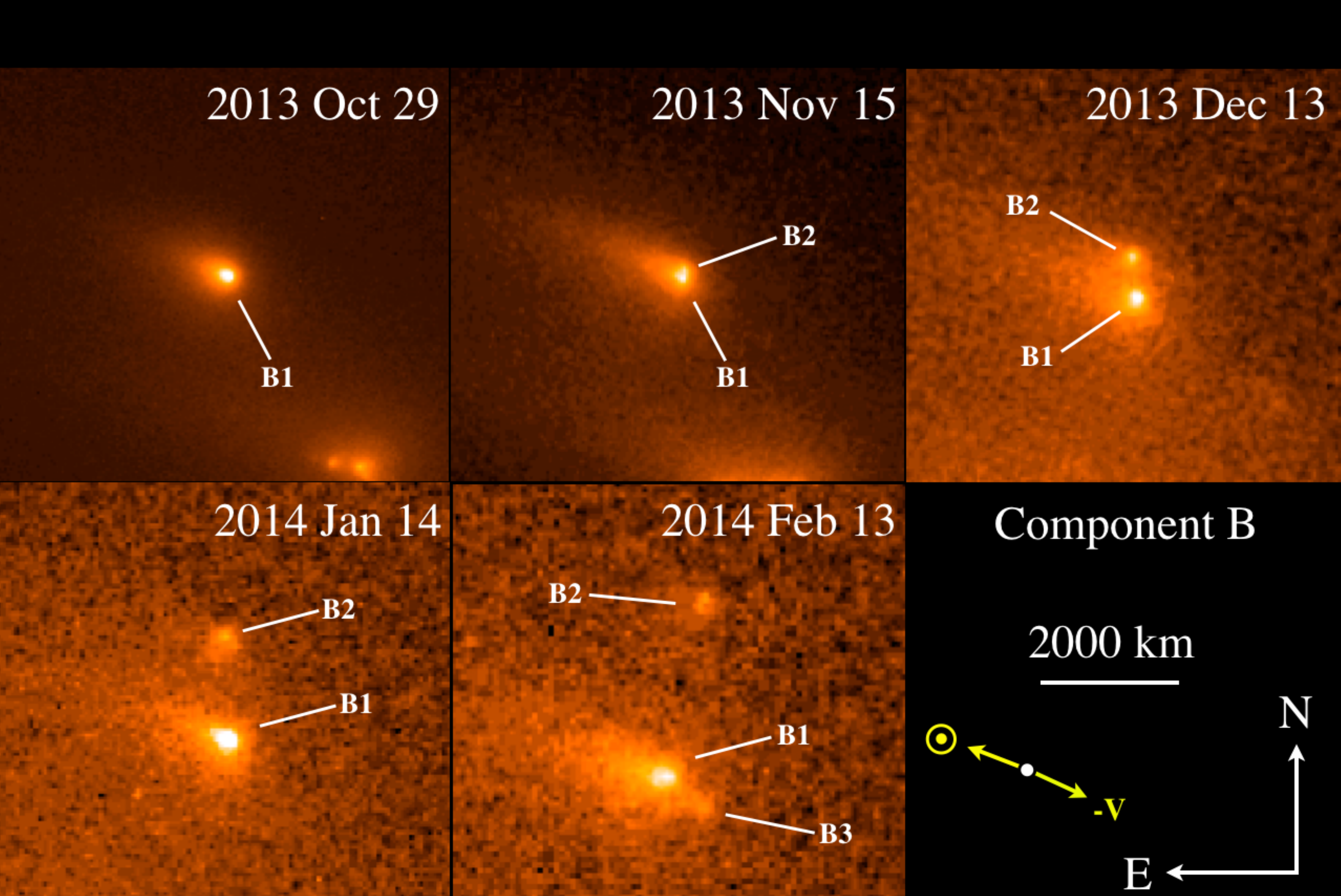}

\caption{Same as Figure (\ref{whole}) but zoomed on component B.   \label{component_B}
} 
\end{center} 
\end{figure}

\begin{figure}
\epsscale{0.85}
\begin{center}
\includegraphics[width=0.8\textwidth, angle =0 ]{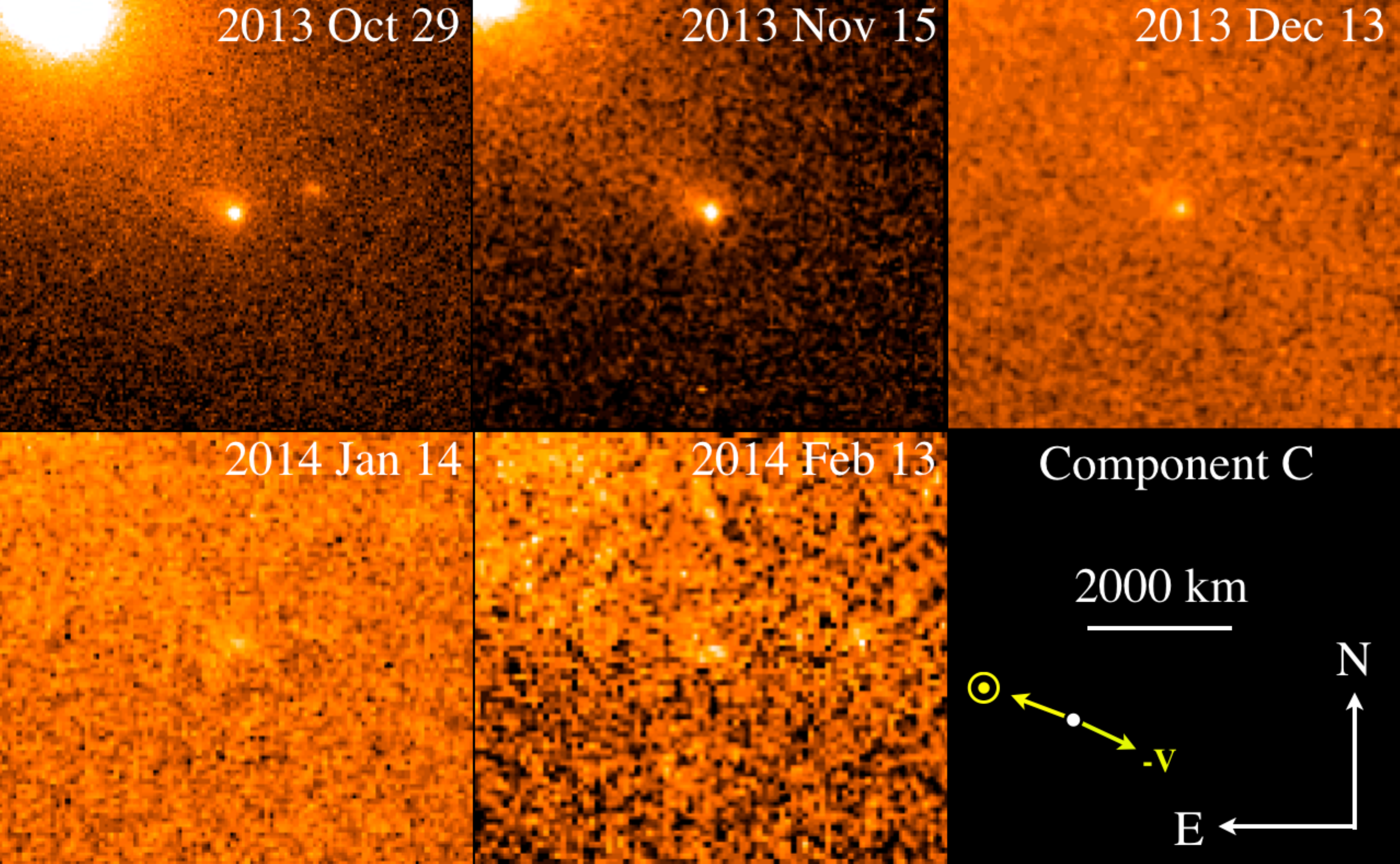}
\includegraphics[width=0.8\textwidth, angle =0 ]{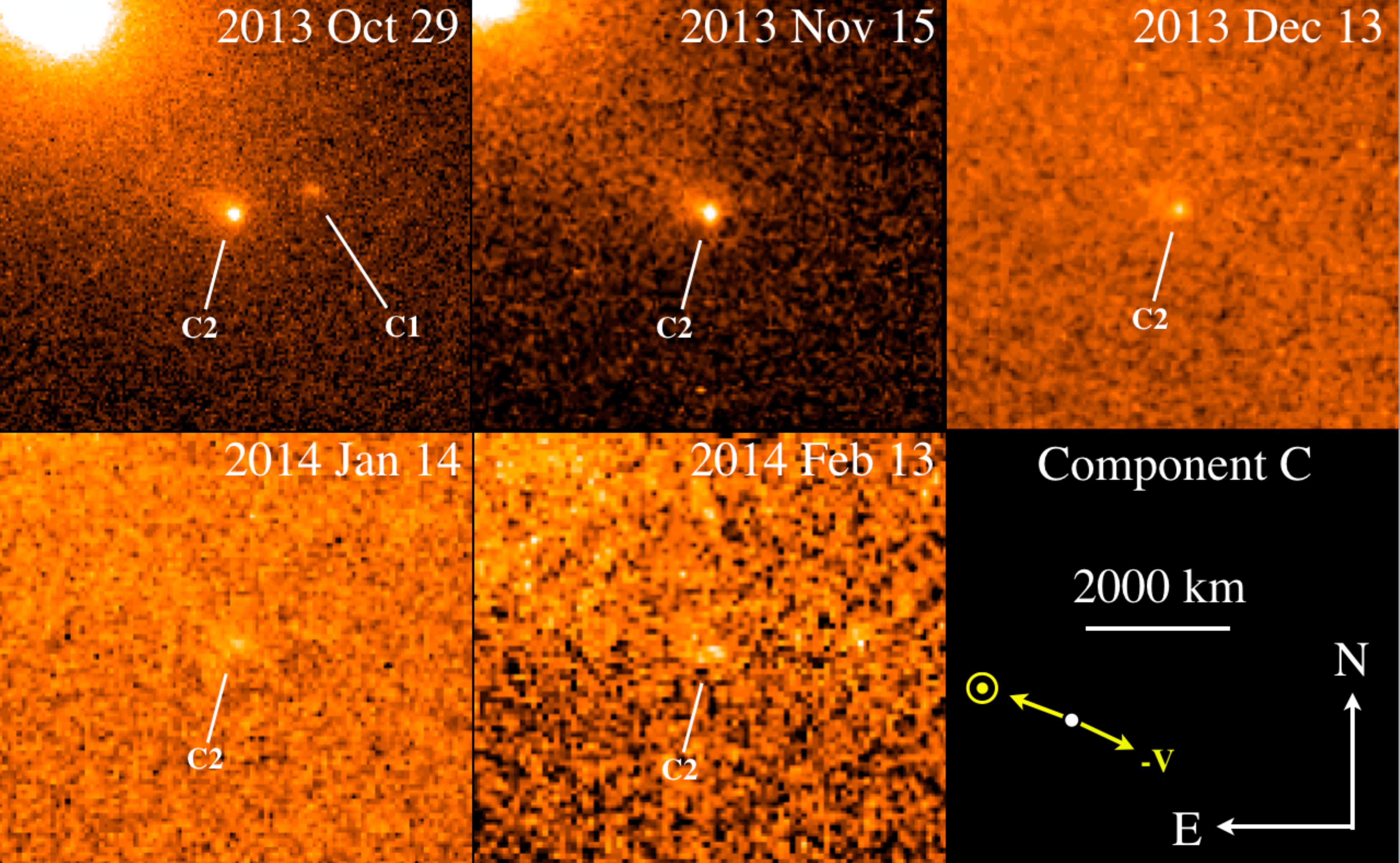}

\caption{Same as Figure (\ref{whole}) but zoomed on component C.   \label{component_C}
} 
\end{center} 
\end{figure}

Figure (\ref{october02}) shows the appearance of R3 in a deep, composite ground-based image from UT 2013 October 02.  In this image, background stars and galaxies have been largely removed by median filter rejection.  The distinctive features of Figure (\ref{october02}) include a broad dust tail to the south west and three objects (labelled A, B and C) embedded in the brightest part of the coma (shown in the inset).  In this first image, the three objects are distributed along a line whose position angle differs from both the antisolar direction and the projected orbit.  These same objects are also apparent in Figure (\ref{october28}), which is another ground-based composite from UT 2013 October 28.  Five epochs of HST data are shown in Figure (\ref{whole}), where A, B and C are clearly resolved into multiple components.  Figures (\ref{component_A}), (\ref{component_B}) and (\ref{component_C}) zoom in to each component to better show their temporal development in the UT 2013 October 29 to 2014 February 13 period.   In all of these images, the position angles of the antisolar vector (marked -$\odot$) and the negative heliocentric velocity vector (marked $-V$) are shown in yellow, and a scale bar is given.  The HST images in Figures (\ref{component_A}), (\ref{component_B}) and (\ref{component_C}) have been scaled by the geocentric distance (from Table \ref{geometry}) in order to facilitate easy comparison at  a fixed, linear scale.

We begin with a brief description of the major components of R3.
 
 \subsection{Component A} 
 Component A is by far the brightest in early October (Figure \ref{october02}) but  fades relative to the other components within a month (Figure \ref{october28}).   High resolution observations on UT October 29 reveal four components in relative motion, increasing to seven by December 13 (Figure \ref{component_A}).  
 
 \subsection{Component B} 
Component B appears single but dusty on UT 2013 October 29, with a small radiation pressure swept tail (position angle 72.5\degr$\pm$1.2\degr, see Figure \ref{component_B}). It develops a nearly perpendicular  elongation (position angle $\sim$350\degr) by UT 2013 November 15 and the UT 2013 December 13 image reveals that a secondary component, B2, has been ejected.  The secondary object on this date possesses its own dust coma and fades progressively from December to the last image on UT 2014 February 13.  The original fragment, B1, brightens in the image from 2014 January 14 and then releases another fragment, B3, in the UT 2014 February 13 image.
 
 \subsection{Component C} 
 Figure (\ref{component_C}) shows that C initially consists of two fragments, one of which (C1) promptly fades to invisibility between UT 2013 October 29 and November 15.  The surviving fragment (C2) itself fades to near invisibility by the end of the observational sequence, with no sign of further fragmentation.

\subsection{Component D} 
Besides the A, B, C components identified in our initial work (Jewitt et al.~2014), we find a fourth object, ``D", in our Keck telescope 5720 s composite integration from UT 2013 October 2  (Figure \ref{october02}).   This faint object lies  down the tail and is widely separated from components A, B and C.  While very faint, it is visible in simultaneous B and R filter sequences recorded using the two independent channels of LRIS. Thus, we are confident that it is not an artifact of the camera or data flattening technique.  Component D is also visible, although barely, in a composite constructed from four 260 s integrations through V and R filters on UT 2013 October 1.  It is not detected in any later data despite its projected position being within the field of view.

\section{FRAGMENT DYNAMICS}
\label{fragdyn}

The individual components of R3 vary  in brightness, position and shape from month-to-month, making it difficult or impossible to link objects through the dataset.  New components appear after the beginning of the HST sequence (e.g.~A3, B2, B3) while others disappear (e.g.~all but two of the seven fragments of component A) before it ends. Some of these changes are caused by the increasing geocentric and heliocentric distances but others are intrinsic to the components.  As a result of the variability, we cannot always reliably link fragments detected in one month to those detected in the next.   Effectively, the monthly cadence of the HST observations is too slow to unambiguously follow the dynamical and photometric evolution of all the fragments.  The reconstruction of the relative motion of fragments is further complicated by their ill-constrained relative masses and the consequent uncertainty of the projected positions of their centers of mass. For these reasons, we have been unable to find meaningful heliocentric orbit solutions to the separate fragments of R3.  Similar difficulties are experienced in the observation of split and fragmented comets where, again, inadequate cadence is the primary culprit (e.g.~Boehnhardt 2004, Jewitt et al.~2016).  

Nevertheless, in some instances, the relationship between separating components is relatively clear.  For example, the images clearly show that components A4, A5 and A6  emanate from A1 (Figure \ref{component_A}).  Fragments B2 and B3 were ejected from B1 (Figure \ref{component_B}).   We measured the positions of those pairs of objects for which the linkage from month to month is relatively clear, in order to estimate the separation velocities in the plane of the sky, and the times of separation.  Our procedure neglects any contribution to the  relative positions from the changing  viewing geometry (Table \ref{geometry}), and also ignores motion parallel to the line of sight.  However, over small intervals these effects should themselves be small.  While the results must  be regarded as only crude estimates,  they are the best we can do in the absence of more densely sampled imaging data taken over a longer temporal arc.    In total, we obtained eight pair-wise measurements of component separation speeds.

\begin{figure}
\epsscale{0.9}
\begin{center}
\includegraphics[width=0.8\textwidth, angle =0 ]{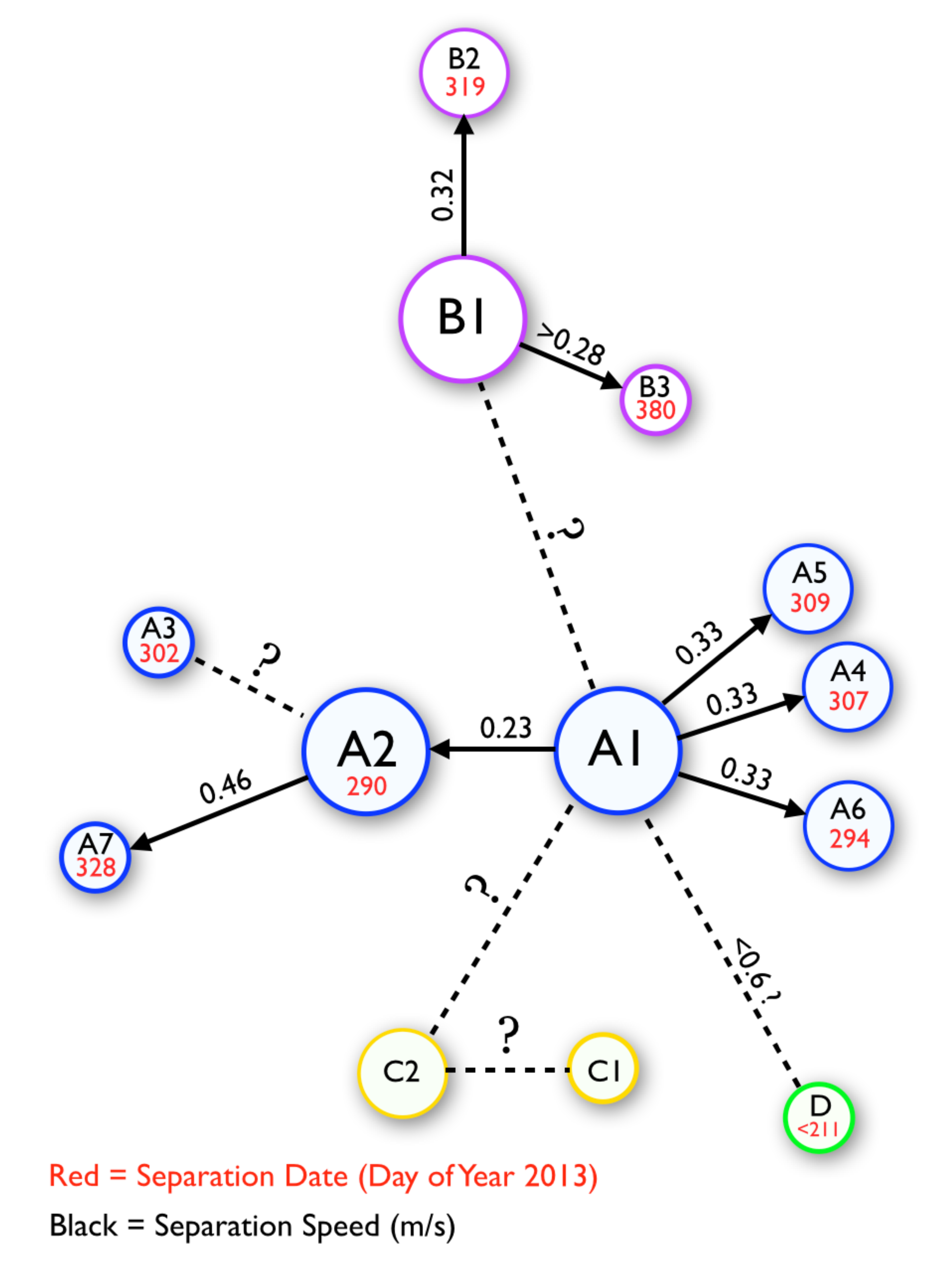}

\caption{Schematic diagram summarizing the relations between the fragments, as discussed in the text.  Black and red numbers indicate, respectively, the estimated separation velocity in the plane of the sky (m s$^{-1}$) and the estimated separation date, expressed as Day of Year number. Colors distinguish the main components, A, B, C and D.  Question marks indicate uncertain relationships. \label{schematic}
} 
\end{center} 
\end{figure}

The results are summarized in Table (\ref{pairs}) and  in Figure (\ref{schematic}), in which the  measured separation velocities are indicated by black numbers and the separation dates are in the Day of Year (DOY)  system (2013 January 1 = 1), shown in red.   Dotted lines indicate uncertain relationships between fragments, for which we have no useful estimates of separation velocity or timing.  In general, this is because the separation occurred sufficiently long before the observations that we cannot establish a reliable connection. From the six fragment-pairs where measurements are possible, the average separation speed is $\Delta v$ = 0.33$\pm$0.03 m s$^{-1}$ (the median is also $\Delta v$ = 0.33 m s$^{-1}$) and the range is only a factor of two, from 0.23 m s$^{-1}$ to 0.46 m s$^{-1}$.   Significantly, there is no evidence for a broader distribution of speeds, for instance, in the form of a halo of faster-moving fragments located outside the central region of the object.  The inferred separation dates vary from DOY 290 (UT 2013 October 17) for the separation of the A1-A2 pair, to DOY 379 (UT 2014 January 15) for B1-B3.  This $\sim$90 day range is again indicative of staggered disintegration, as concluded below from the position angles of the dust tails.

The observations of component D have very poor temporal coverage (observations only on two consecutive days). The separation between components A1 and D was $\ell$ = 36.6\arcsec$\pm$0.2\arcsec~on UT 2013 October 01 and 37.0\arcsec$\pm$0.1\arcsec~on October 02, corresponding to about 33,000 km projected to the distance of R3.  The difference, 0.4\arcsec$\pm$0.2\arcsec~is statistically consistent with zero (Table \ref{pairs}), and we interpret the position measurements as setting a 3$\sigma$ upper limit to the motion of 0.6\arcsec~(540 km) in 1 day.  The resulting limit to the sky-plane speed is $v \le$ 6 m s$^{-1}$ and the travel time $t_D \sim \ell/v \ge$ 5.5$\times$10$^{6}$ s ($\gtrsim$ 2 months, corresponding to DOY $\le$ 211).  If we instead assume that D was released from the parent body with negligible initial velocity and experienced constant acceleration away from the source, then we derive acceleration $\lesssim$ 5$\times$10$^{-7}$ m s$^{-2}$ and travel time $t_D$ $\gtrsim$ 1.5$\times$10$^{7}$ s (6 months, or DOY $\lesssim$ 93).  The truth is likely to lie between these values.  Again, the A1-D separation time is distinctly different from the others listed in Table (\ref{pairs}), showing staggered disintegration.

\section{DUST MORPHOLOGY}
The dust morphology changes dramatically as a function of time (e.g.~compare Figures \ref{october02} and \ref{october28}).  The ground based images from October 02 and 28 show extended emission far beyond the domain occupied by components A, B and C, but do not reveal useful information on the sub-arcsecond scale.   The finer resolution of the HST, on the other hand, reveals mini-tails in association with many of the sub-components of A, B and C (see Figures (\ref{component_A}), (\ref{component_B}) and (\ref{component_C})).  These mini-tails are presumably present, but cannot be resolved, in the ground-based data.  To study  changes in the dust tail, we measured the position angles of the dust tails of R3 on each date of observation.   While measurements from the Keck October 02 data were straightforward, we were forced to smooth the less deep Magellan composite from October 28 by convolution with a Gaussian function having a standard deviation of 5 pixels (1.0\arcsec) in order to improve the signal-to-noise ratio.  Likewise, the tiny pixels of the HST images (0.04\arcsec~on a side, compared to 0.135\arcsec~for Keck and 0.2\arcsec~for Magellan) were Gaussian smoothed in order to improve the detectability of faint extended emission.

The tail position angles were determined by least-squares fitting a linear relation to the mid-line of each tail, determined visually.  The position angle measurements are summarized in Table (\ref{positionangle}). Our ability to locate the mid-line was limited by some combination of contamination by trailed background objects, noise in the data, scattered light from bright, nearby  stars and blending of tails emitted by separate components of R3.  In  view of these many complicated and hard-to-model influences, we take the  uncertainty of the position angle calculated  from the least-squares fit (typically $\pm$ 0.3\degr~for the west tails and $\pm$ 1\degr~for the east tails) as lower limits to the true uncertainties.

\begin{figure}
\epsscale{0.95}
\begin{center}
\includegraphics[width=0.95\textwidth, angle =0 ]{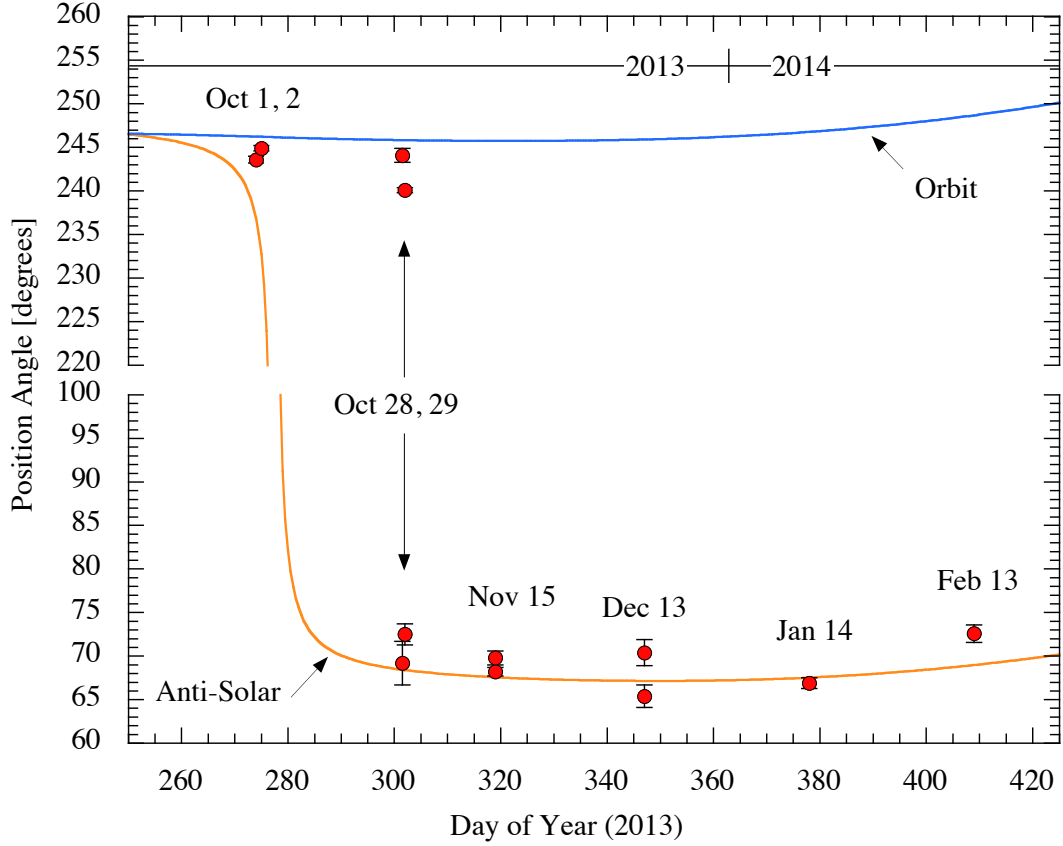}

\caption{Dust tail position angles (red circles) measured as a function of Day of Year (1 = UT 2013 January 01).   The solid lines show, respectively, the position angle of the negative heliocentric velocity vector (marked ``Orbit'', blue line) and the anti-solar direction (orange line). Position angles in the range 100\degr~to 220\degr~are omitted for clarity of presentation. Plotted position angle uncertainties are statistical only, and neglect poorly quantified systematic errors likely to be several times larger. \label{angles_plot}
} 
\end{center} 
\end{figure}

Figure (\ref{angles_plot}) shows the resulting position angles measured as a function of the date of observation.   Also plotted in the figure are lines showing the evolution of the anti-solar position angle projected into the plane of the sky (orange line) and of the negative heliocentric velocity vector (blue line).  The Figure shows that the position angles fall either close to the anti-solar direction or near to the projected orbit.     The widest separation occurs in observations from  October 28 and 29, when the projected anti-solar and  orbit directions are almost 180\degr~apart, and two tails are observed.  The tail to the east (position angles $\sim$70\degr) consists of particles small enough to be swept by solar radiation pressure while the tail to the west (position angles $\sim$242\degr) consists of large, slow-moving particles insensitive to solar radiation pressure and concentrated near the orbit of the R3 precursor.  After October, the large particle tail is no longer apparent but a series of tails aligned with the anti-solar direction emanate from various components of A and B.  The measured position angles of the anti-solar tails (Figure (\ref{angles_plot})) are generally consistent with synchrones (Finson \& Probstein, 1968) of dust emitted after mid-September and continuing to at least mid-January. We were not able to associate a particular dust ejection date to a particular fragment, reflecting either continued activity or the large uncertainties of the position angle measurements, or both. Protracted  dust production from the fragments is independently indicated by the photometry (Section \ref{individuals}). Even without numerical modeling, Figure (\ref{angles_plot}) shows that small dust particles continued to be emitted from the individual components of R3 over a protracted period (130 days or more), arguing against a collisional origin for the disruption.  

\section{PHOTOMETRY}

The complex and evolving optical morphology of R3 presents many challenges to the extraction of useful photometry.  In the interests of simplicity, we elected to use aperture photometry, where possible, to measure the brightnesses of the various fragments of this object.  We used a circular aperture of fixed projected  radius $\ell$ = 195 km (0.4\arcsec~at 1.34 AU, the minimum distance of R3 in the HST observations (Table \ref{geometry})). Background subtraction was obtained from a contiguous annulus of outer radius 390 km, to obtain the measurements summarized in Table (\ref{photometry}).  We also obtained photometry capturing all of the fragments within a single, large aperture of radius 8000 km, labeled ``diffuse'' in the Table.

The apparent magnitudes were converted into absolute magnitudes using the inverse square law

\begin{equation}
H_V= V - 5\log(r_H\Delta) - 2.5\log_{10}(\Phi(\alpha))
\label{absolute}
\end{equation}

\noindent where $\Phi(\alpha)$ is the phase function, equal to the ratio of the scattered light at  phase angle $\alpha$ to that at $\alpha$ = 0\degr.   We assumed the phase function formalism of Bowell et al.~(1989) with parameter $g$ = 0.15, as appropriate for a C-type object (Jewitt et al.~2014).  Although, the phase function of R3 is unmeasured, the uncertainty introduced by this assumption is modest, given that the largest phase angle at which R3 was observed is only 24\degr~(Table \ref{geometry}).  For example, the correction and the derived $H$ values would be at most $\sim$0.13 magnitudes fainter if we had instead assumed  the phase function of an S-type asteroid ($g$ = 0.25).  Absolute magnitudes are given in Table (\ref{photometry}).

The absolute magnitudes are in-turn related to the scattering cross-section by

\begin{equation}
C_e = \frac{1.5\times10^6}{p_V} ~10^{-0.4 H_V}
\label{area}
\end{equation}

\noindent where $p_V$ is the geometric albedo, here assumed to be 0.05 (as in Jewitt et al.~2014).  The derived scattering cross-sections are also given in Table (\ref{photometry}). 

The large aperture photometry (labeled ``diffuse'' in Table \ref{photometry}) gives a minimum total cross-section $C_e$ = 20.4 km$^2$, with an uncertainty that is dominated by the unmeasured albedo, $p_V$ (a 50\% error in $p_V$ corresponds to a 50\% error in $C_e$).  This cross-section is dominated by dust scattering and sets a strong upper limit to the size of the parent body which disintegrated into the multiple components of R3.   The radius of the equal-area circle is $r_e = (C_e/\pi)^{1/2}$, which gives $r_e$ = 2.5 km.  A more stringent constraint on the precursor body size may be obtained by taking the smallest determination of $C_e$ on a component-by-component basis, again from Table (\ref{photometry}).  Then, the effective radius of the precursor body is given by $r_e = (\Sigma C_e/\pi)^{1/2}$,  where $\Sigma C_e$ is the sum of the minimum cross-sections of each fragment from the Table.  We find $\Sigma C_e$ = 0.53 km$^2$, giving $r_e$ = 0.4 km.   This is still an overestimate of the cross-section and radius because of dust contamination remaining in the photometry apertures used in Table (\ref{photometry}).  

Photometry of component D in the Keck October 01 composite data gives a mean R-band magnitude $m_R$ = 24.9$\pm$0.2, while on October 02 we find $m_R$ = 24.8$\pm$0.1.  Both measurements were taken within a 6\arcsec~radius projected aperture, with background subtraction from a contiguous annulus having outer radius 7.3\arcsec. Absolute magnitude of D is $H_V$ = 22.45 which, with albedo $p_V$ = 0.05, implies an effective radius $r_e \sim$ 100 m.  The object is too faint for its surface brightness profile to be meaningfully measured and it is possible that some of the measured brightness is contributed by dust emitted from a smaller source.  Therefore, we can only conclude that the data suggest the presence of a solid body not larger than $r_e \sim$ 100 m.

\subsection{The Diffuse Debris Envelope}
\begin{figure}
\epsscale{0.85}
\begin{center}
\includegraphics[width=0.95\textwidth, angle =0 ]{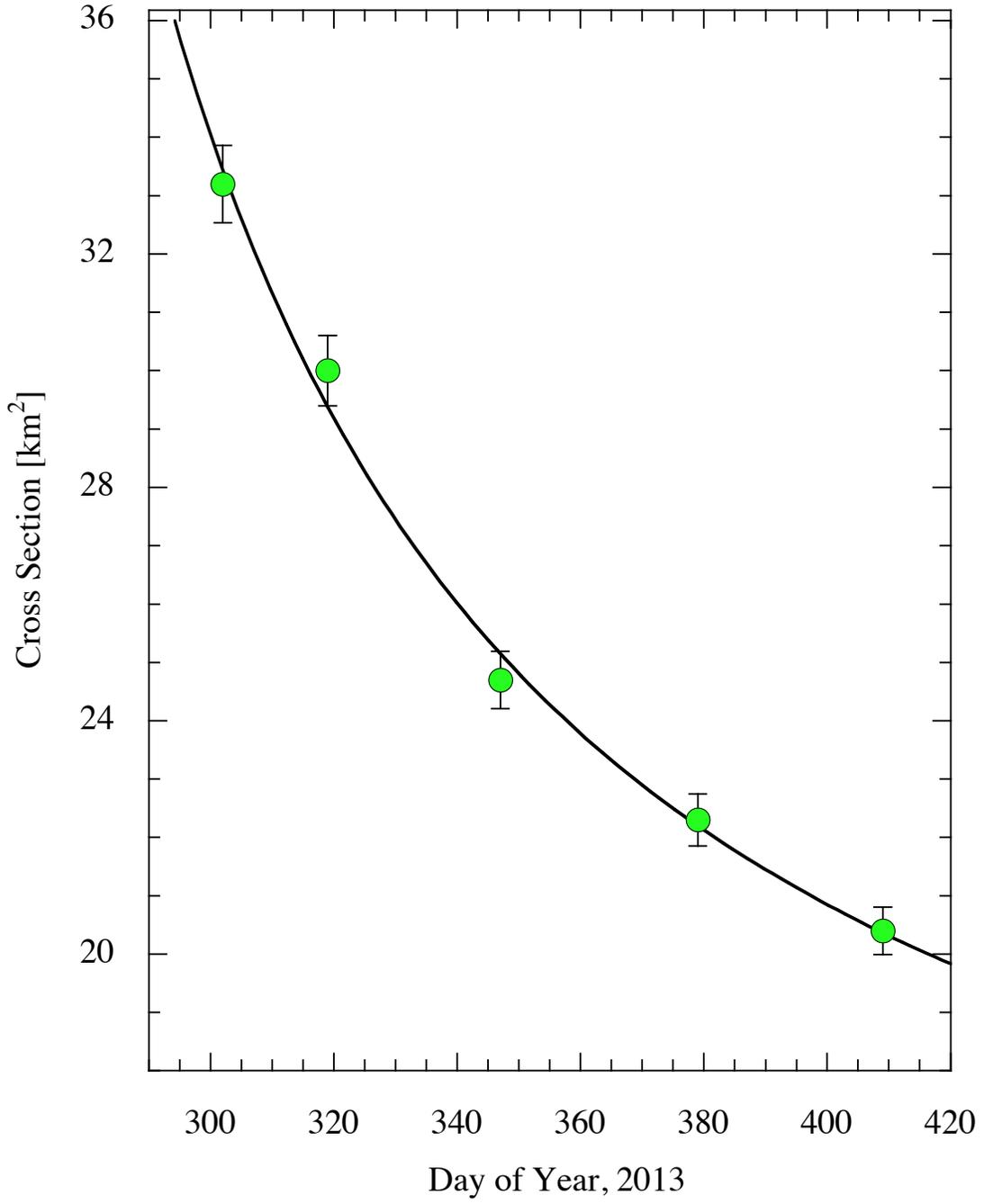}

\caption{Photometry in an aperture of fixed projected radius 8000 km (Table (\ref{photometry})), converted to scattering cross-section (Equation (\ref{area})) and fitted using Equation (\ref{fitter}).    Error bars of $\pm$2\% are shown. \label{modelfits}
} 
\end{center} 
\end{figure}

Table (\ref{photometry}) shows that the total cross-section of the diffuse debris in R3 is much larger than the cross-sections of the discrete components, even when the latter are added together.  We  focus our attention on  large-aperture photometry measured within a circle of fixed linear radius 8000 km, scaled to the distance of R3.  This aperture is large enough to capture the bulk of the light scattered by dust but small enough to avoid photometry problems associated with the sky background distant from the optocenter of R3.  It also excludes the very faint component D.  The measurements are plotted in Figure (\ref{modelfits}), where a monotonically decreasing cross-section is apparent.

To interpret Figure (\ref{modelfits}) we consider a model in which the ejected dust follows a size distribution such that the number of particles with radii in the range $a$ to $a + da$ is $n(a)da = \Gamma a^{-q} da$, where $\Gamma$ and $q$ are constants.  The size distribution is assumed to range from minimum radius $a_{min}$ to maximum $a_{max}$. We assume that the particles are released with zero initial speed at time $T_0$ and then accelerated by solar radiation pressure, of magnitude $g$.  The radiation pressure acceleration is expressed as a multiple of the gravitational acceleration towards the Sun, or

\begin{equation}
g = \beta \frac{G M_{\odot}}{r_H^2}
\label{alpha}
\end{equation}

\noindent in which $G$ is the gravitational constant, $M_{\odot}$ is the mass of the Sun and $r_H$ is the heliocentric distance. The quantity $\beta$ is the dimensionless radiation pressure efficiency factor (e.g.~Bohren and Huffman 1983) which, to a first approximation, is given by $\beta = 1/a_{\mu m}$, where $a_{\mu m}$ is the particle radius expressed in microns.    Since the heliocentric distance changed little (Table \ref{geometry}) we assume a fixed value $r_H$ = 2.35 AU in all that follows.  

The distance, $\ell$, traveled in time, $t$,  by a particle under fixed acceleration, $\alpha$, is just $\ell = 1/2 \alpha t^2$.  We set $\ell$ = 8000 km, corresponding to the linear radius of the projected photometry aperture, recognizing that this gives a lower limit to the true distance traveled because of the effects of projection.  Substituting Equation (\ref{alpha}) and $\beta = 1/a_{\mu m}$, we solve for the radius of the dust particle that reaches the edge of the aperture at time  $t$, 

\begin{equation}
a_c(t) = \frac{G M_{\odot}}{2 \ell r_H^2} (t - T_0)^2,
\label{acrit}
\end{equation}

\noindent where $T_0$ is the time of their release from the nucleus and $t-T_0 \ge 0$ is the time of flight.  
Equation (\ref{acrit}) gives the radius of the smallest particle, in microns, which has not been swept from the photometry aperture by radiation pressure in the time $t - T_0$.  

The total cross-section of the dust remaining in the aperture as a function of elapsed time is then given by

\begin{equation}
C_d(t) =  \pi \Gamma \int_{a_{c}(t)}^{a_{max}} a^{2 - q} da.
\end{equation}

\noindent provided $a_{max} \ge a_c(t)$. For $q \ne 3$, 

\begin{equation}
C_d(t) =  \frac{\pi \Gamma}{3-q}  \left[a^{3 - q}\right]_{a_{c}(t)}^{a_{max}} 
\end{equation}

\noindent which, with $a_{max} \gg a_c$ and substituting Equation (\ref{acrit}) becomes

\begin{equation}
C_d(t) =  \frac{\pi \Gamma}{q-3} \left(\frac{G M_{\odot}}{2 \ell r_H^2}\right)^{3 - q} (t- T_0)^{6-2q}
\label{coft}
\end{equation}

\noindent At very small flight times, $(t - T_0)$, we anticipate that Equation (\ref{coft}) should fail because the smallest optically important particles in the size distribution will not have had time to reach the edge of the photometry aperture.  However, substitution of $a_c$ = 0.1 $\mu m$  into Equation (\ref{acrit}) shows that this initial period in which Equation (\ref{coft}) is inapplicable is $t - T_0 \lesssim$ 3 hr, which is negligible.  The relation will also fail at early times because the optical depth can exceed unity (again, only for a timescale $\sim$1 hour); we ignore this short-lived and unobserved regime.  

Measurements of the dust released from fragmented active asteroids P/2010 A2 ($q$ = 3.3$\pm$0.2, Jewitt et al.~2010) and P/2012 F5 (Gibbs) ($q$ = 3.7$\pm$0.1, Moreno et al.~2012) and from  fragmenting comet 332P/Ikeya-Murakami ($q$ = 3.6$\pm$0.6, Jewitt et al.~2016) are  consistent with $q$ = 3.5. If this value applies to R3, then Equation (\ref{coft}), gives an inverse dependence of cross-section on time, . 

\begin{equation}
C_d(t) =  \frac{2 \pi \Gamma} {(t- T_0)} \left(\frac{G M_{\odot}}{2 \ell r_H^2}\right)^{-1/2}
\end{equation}

Writing the total scattering cross section, $C(t)$, as the sum of the cross-sections of the nucleus, $C_0$, and the dust, or $C(t) = C_0 + C_d(t)$, we next fitted the fixed-aperture photometry by least-squares using 

\begin{equation}
C(t) =  C_0 + \frac{K}{(t-T_0)}
\label{fitter}
\end{equation}

\noindent where $C_0$, K $ = 2\pi\Gamma (2 \ell r_H^2/G M_{\odot})^{1/2}$ and $T_0$ are constants.

Evidently, Figure (\ref{modelfits}) shows that Equation (\ref{fitter}) presents a rather good fit to the data.   The best-fit is given by  $C_0$ = 11$\pm$3 km$^2$, K = 1730$\pm$370 km$^2$ days, and $T_0$ = 225$\pm$24 days, corresponding to UT 2013  August 13$\pm$24.  To estimate the  uncertainties of the fit, we assumed that each of the individual cross-section measurements is uncertain to within $\pm$2\%.  The initiation times in mid-August, as they should, pre-date the discovery of R3 on UT 2013 September 13 (Hill et al.~2013), and are compatible with ground-based images showing components A, B and C as widely separated on UT 2013 October 01 and 02 (Jewitt et al.~2014).

We note that $C_0$, which measures the cross-section of the nucleus and any material too large to be accelerated beyond the 8000 km photometry aperture in the period of observations, is formally significant at the 3$\sigma$ level. We thus set an upper limit to the effective radius of any precursor solid body $r_0 = (C_0/\pi)^{1/2}$ = 1.9 km.

In our model, the fading  is a result of  the preferential escape of small dust particles from the photometry aperture, with the radius of the smallest particle contributing to the cross-section given by Equation (\ref{acrit}).  Substituting $T_0$ = 225, we find $a_{c} \sim$ 3 mm at $t$ =302 (UT 2013 October 29) rising to $a_{c} \sim$ 20 mm by $t$ =409 (UT 2014 February 14).   The diffuse debris envelope consists  of large particles having a combined mass given by

\begin{equation}
M = \frac{4\pi \rho \Gamma}{3}\int_{a_{c}(t)}^{a_{max}} a^{3-q} da. 
\label{mass2}
\end{equation}

\noindent  We assume $a_{max} \gg a_{c}(t)$, $q$ = 3.5 and use Equation (\ref{coft}) to eliminate $\Gamma$ from Equation (\ref{mass2}), finding

\begin{equation}
M = \frac{4}{3}\rho C_d(t) (a_{c} a_{max})^{1/2}.  
\label{mass}
\end{equation}

\noindent The maximum dust particle size is unknown.  To consider an extreme, we set $a_{max}$ = 200 m, which is comparable to the sizes of the largest components observed in R3 (Table \ref{photometry}).  Then, for $t$ = 302 days, with $a_{c}$ = 3 mm and $C_d = C - C_0$ = 22 km$^2$ and $\rho$ = 10$^3$ kg m$^{-3}$, Equation (\ref{mass}) gives the characteristic mass $M$ = 2.3$\times$10$^{10}$ kg.  While this model is clearly simplistic, it shows that the upper limit (given that we picked $a_{max}$ to be extreme) to the mass of the diffuse debris envelope is equivalent to an equal-density sphere of radius $\sim$170 m. 


\subsection{Individual Components}
\label{individuals}
We find that, unlike with the diffuse debris envelope, Equation (\ref{fitter}) does not provide a good description of the photometry of the individual components of R3. Presumably, this is because, as the images suggest, these objects continue to disintegrate in the period of observations.  We illustrate this using component B, which splits within the window of the HST observations (Figure \ref{component_B}).  A fit of Equation (\ref{fitter}) to the October, November and December measurements of B gives $C_0$ = -0.013$\pm$ 0.065 km$^2$,  K = 17.4$\pm$4.0 km$^2$ days and $T_0$ = 290$\pm$2 days (UT 2013 October 17$\pm$2).  The 3$\sigma$ limit to $C_0 \le 0.20$ km$^2$, compares with the smallest cross-section measured photometrically, namely $C$ = 0.17 km$^2$ (Table \ref{photometry}).  A fit to the three measurements in which B2 is clearly separated from B1 (on 2013 December, 2014 January and February) gives $C_0$ = 0.049$\pm$0.029, K = 4.29$\pm$2.75 and $T_0$ = 323$\pm$22 (UT 2013 November 19$\pm$22).  Clearly, these solutions are of limited significance, given that they are fits of three parameters to three data-points, each. However, $T_0$ for B2 is consistent with the astrometric data, which show B2 separating from B1 in mid-November (Table \ref{pairs}).

We examined the individual images from HST to search for  photometric variations occurring within a single orbit (i.e.~on timescales $\lesssim$ 1 hour).  For this purpose we employed the smallest reasonable photometry aperture (0.2\arcsec~in radius) with background subtraction from a concentric annulus having inner and outer radii 0.2\arcsec~and 0.4\arcsec, respectively.  This small aperture minimizes the contribution from the background coma and reduces the deleterious effects of near-nucleus cosmic rays.  In a few instances we digitally removed cosmic rays from the data by interpolation but, in general, we simply omitted  data contaminated by cosmic rays from further consideration.  

\begin{figure}
\epsscale{0.95}
\begin{center}
\includegraphics[width=0.95\textwidth, angle =0 ]{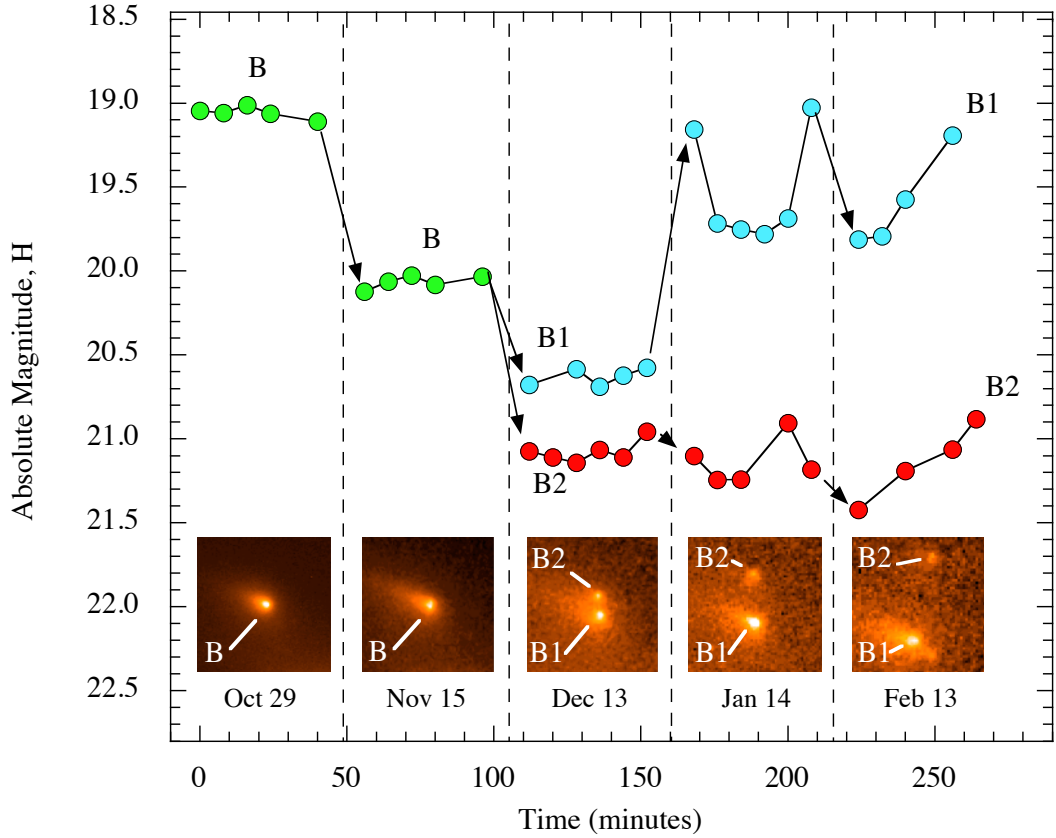}

\caption{Lightcurve of component B as a function of time in HST observations from 2013 October to 2014 February.  Component B (green circles) is resolved into B1 (blue circles) and B2 (red circles) in  observations taken after 2013 November, as shown in the image panels included in the figure.  A third component, B3 (c.f.~Figure \ref{component_B}) emerges from B1 in the February 13 data but is not plotted here.  \label{Figure_B}
} 
\end{center} 
\end{figure}

The most interesting measurements are those of fragment B, and are plotted in Figure (\ref{Figure_B}). In order to present the widely-spaced data in a single figure we have compressed the time axis such that a scale of hours applies to each epoch of observation but with an arbitrary time offset between epochs, denoted in the figure by vertical dashed lines.  As noted earlier, within each HST orbit we obtained six images of R3.  The absolute magnitude (Equation \ref{absolute})  is shown in order to remove most of the effects of the changing observing geometry.  

Figure (\ref{Figure_B}) confirms that the absolute magnitude of component B varies in conjunction with morphological changes (shown in five small image panels in the figure).  Component B is uniformly bright ($H_V$ = 19.1, $C_e$ = 0.71 km$^2$) on October 29 but has faded considerably ($H_V$ = 20.1, $C_e$ = 0.28 km$^2$) by November 15, at which time the core region of the object has taken an elongated appearance owing to the barely resolved separation of components B1 and B2.  By December 13, these components are fully resolved. The fainter component, B2, remains photometrically steady ($H_V \sim$  21.1, $C_e \sim$ 0.11 km$^2$) compared to B1, which flares from $H_V$ = 20.6 ($C_e$ = 0.17 km$^2$) to $H_V$ = 19.5 ($C_e$ = 0.47 km$^2$) on January 14 and to 19.6 ($C_e$ = 0.44 km$^2$) on  February 13.  The fluctuations in data from January 14 and February 13 are an order of magnitude larger than expected from the statistical uncertainties of the photometry ($\pm$0.02 and $\pm$0.05 magnitudes, respectively) and are presumed to be real.  However, they do not appear to be caused by nucleus rotation since the time-resolved photometry presents no convincing evidence that the variations are cyclic.

The initially bright state of component B (October 29) precedes its separation into B1 and B2 (on UT 2013 November 15$\pm$10, according to Table \ref{pairs}), two components that were fully resolved by December 13.  Likewise, the subsequent brightening of B1 in January and February may be associated with the separation of B3.  This pattern, in which separation of components is preceded by a brightness surge, has been reported previously in other split comets.  The photometric flaring (Ishiguro et al.~2014) and fragmentation (Jewitt et al.~2016) of 332P/Ikeya-Murakami constitutes an outstanding recent example, while a list of cases is compiled in Boehnhardt (2004).   

The distinctive mini-comet morphology of the components in Figure (\ref{component_B}) shows the influence of solar radiation-pressure.  In particular, the length of the sunward ``nose'', $\ell_u$, corresponds to the distance at which particles ejected towards the Sun are turned back by radiation pressure.  Measurements show that the sunward surface brightness falls to 1/2 of its peak value at angular distance $\ell_u$ = 0.5\arcsec~from the photocenter of component B in the November image. The corresponding linear distance, $\ell_u$ =500 km, gives an estimate of the turnaround distance for the optically important particles, again neglecting the effects of projection. If we again write the acceleration as a multiple, $\beta$, of the gravitational acceleration towards the Sun,  $g_{\odot}(r_H)$, then

\begin{equation}
\beta = \frac{u^2}{2 g_{\odot}(r_H) \ell_u}
\label{beta}
\end{equation}

\noindent in which $u$ is the speed of the ejected particles.  We approximate the latter by $u = p /\tau$, where $p$ is the projected radius of the photometry aperture and $\tau$ is the timescale over which particles escape the aperture, for which we use the empirical fading timescale.  Then,

\begin{equation}
\beta = \left(\frac{r_H^2}{2G M_{\odot} \ell_u}\right)\left(\frac{p}{\tau}\right)^2 
\label{beta2}
\end{equation}

\noindent where $r_H$ is expressed in meters.  For the 0.2\arcsec~radius aperture and geocentric distance $\Delta$ = 1.4 AU, as in 2013 November (see Table \ref{geometry}), we have $p$ = 200 km.  From Figure (\ref{Figure_B}), component B fades by a factor $\sim$2 between October 29 and November 15, giving $\tau \le$ 17 days (1.5$\times$10$^6$ s).  
With $G$ = 6.67$\times$10$^{-11}$ N kg$^{-2}$ m$^2$, $M_{\odot}$ = 2$\times$10$^{30}$ kg, $r_H$ from Table (\ref{geometry}), substitution into Equation (\ref{beta2}) gives $\beta \ge$ 10$^{-5}$.  The relation between $\beta$ and particle size depends upon unknowns including the density and the porosity of the material.  Assuming density $\rho$ = 10$^3$ kg m$^{-3}$, $\beta \ge$ 10$^{-5}$ corresponds roughly to an effective particle radius $a \le$ 0.1 m (Bohren and Huffmann 1983). 

The inferred dust ejection speed, computed from $u = p/\tau$, is $u \ge$ 0.13 m s$^{-1}$.  It is worth noting that this modest dust velocity is of the same order of magnitude as the velocity dispersion of the major ($\sim$100 m scale) components of R3, for which we found $\Delta v$ = 0.3 m s$^{-1}$ (Section \ref{fragdyn}).  A size-independent ejection velocity is  inconsistent with comet-like gas-drag acceleration (which produces $u \propto a^{-1/2}$) but is expected from a rotational breakup in which the characteristic speed is the gravitational escape speed from the disrupting body.

\begin{figure}
\epsscale{0.95}
\begin{center}
\includegraphics[width=0.95\textwidth, angle =0 ]{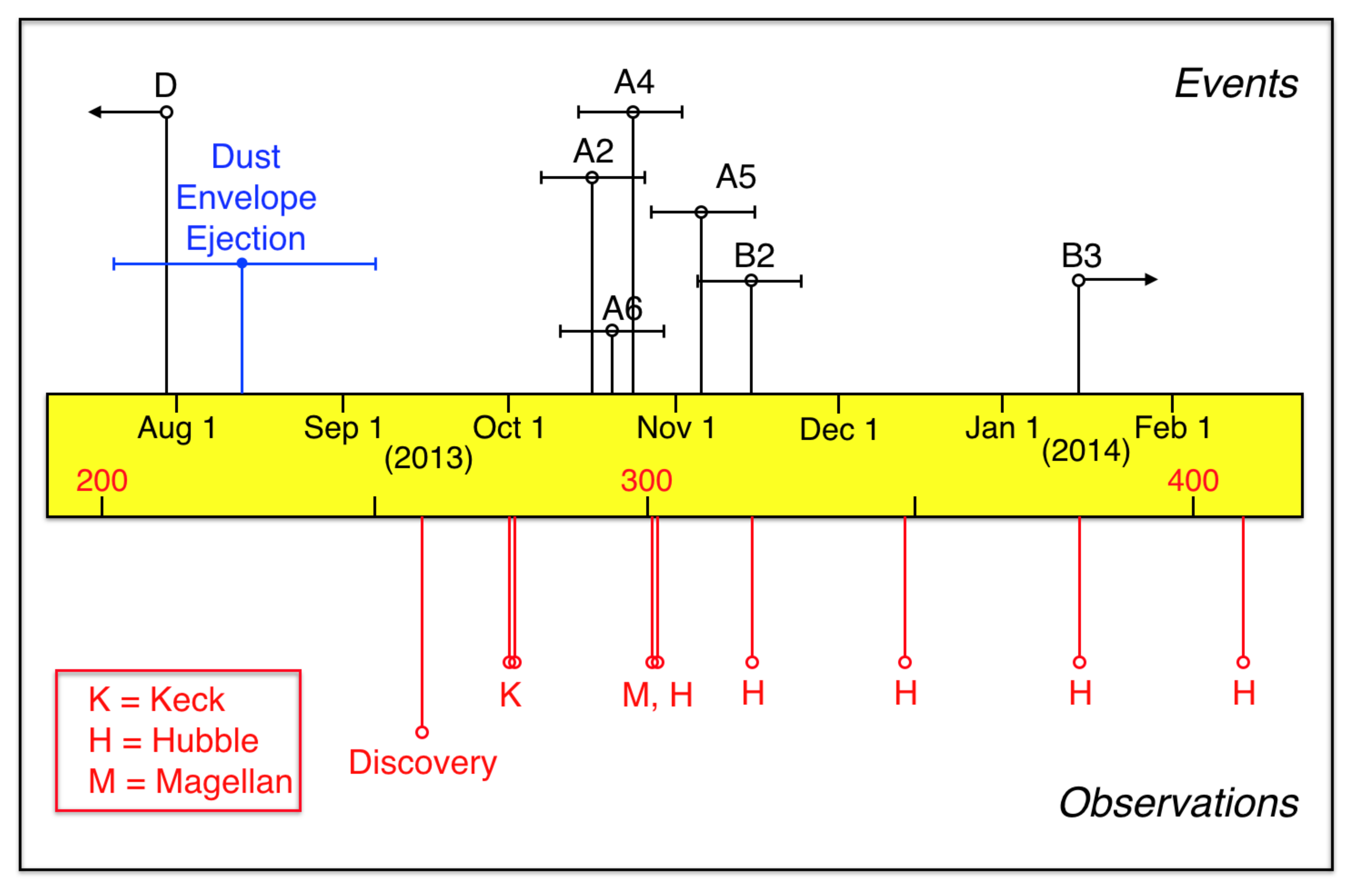}

\caption{Timeline for R3, showing calendar dates in black and Day of Year in red.  Events in R3 are shown above the yellow timeline box while the dates of observations discussed in the text are indicated below it. The late-stage non-detection observations in 2015 are not shown.  \label{timeline}
} 
\end{center} 
\end{figure}

Figure (\ref{timeline}) presents the times of various events in R3 deduced from the images.  It shows that the estimated time of the emplacement of the diffuse debris envelope overlaps with the limit on the time of ejection of component D, perhaps that these two events are related.  A cluster of fragment separation times appears between mid-October and mid-November, followed by a quiet period until the separation of B3 from B1 in mid-January.  Most importantly, the Figure shows that the fragmentation of R3 is spread over a time interval $\sim$6 months.



 \section{LATE-STAGE NON-DETECTIONS}
 The solar elongation of R3 fell below 50\degr~between 2014 March 11 and September 04, rendering HST observations impossible.  After conjunction, we observed again using the HST and WFC3, on a range of dates from UT 2014 September 29 to 2015 May 26 (Table \ref{geometry}).  R3 was not detected in any of these post-conjunction observations.  The 5$\sigma$ limiting magnitude on a point-source target, obtained by combining all the images taken within a given orbit, is $V$ = 28.0.  This limit, corrected for the observing geometry using Equation (\ref{absolute}), gives a limit on the absolute magnitude of any fragments within the field of view varying from $H_V >$ 21.7 on UT 2015 May 26 to $H_V >$ 23.5 on UT 2015 January 17, and these limits correspond to equivalent  areas $C_e <$ 0.06 km$^2$ and $C_e <$ 0.01 km$^2$, respectively,  by Equation (\ref{area}),  again with $p_V$ = 0.05.   Equal-area circles have radii $r_e <$ 0.14 km and $r_e <$ 0.06 km, respectively.  It is tempting to assert that these are stringent limits on the dimensions of surviving fragments of R3, but possible ephemeris uncertainties  leave residual doubt about this interpretation.  The formal 3$\sigma$ uncertainties in the position of R3 in the post-conjunction observations vary from about $\pm$40\arcsec~(2014 September 29) to about $\pm$70\arcsec~(2015 January 17), according to the JPL Horizons ephemeris.  These values are small compared to the 123\arcsec$\times$136\arcsec~WFC3 field of view but the formal uncertainties may not reflect the actual uncertainties inherent in astrometric measurements of a morphologically complex, evolving target like R3.  In addition to being morphologically complex, R3 was diffuse and faint, limiting the arc of astrometric observations employed in the Horizons orbit solution to only 107 days (UT 2013 October 29 to 2014 February 13).   As a result, we are not confident that the formal astrometric uncertainty from Horizons is a good measure of the true uncertainty, and so we cannot be certain that R3 was within the HST field of view.
 
To try to test the possibility that fragments of R3 might be outside the HST field of view in our post-conjunction data, we used the wider coverage (unvignetted field is about 360\arcsec~square) of the Keck telescope, obtaining observations on UT 2015 February 17 and 2015 December 08.    The more sensitive observations were obtained on the former date and we discuss them exclusively.  We placed the expected location of R3 near the center of the field of view, but avoiding the gap between CCDs, and obtained six consecutive R-band images, each of 600 s duration and, simultaneously, six integrations of 640 s through the B-band filter.  The images were tracked non-sidereally, using the JPL Horizons ephemeris for  ``2013 R3-B'' in seeing of $\sim$0.8\arcsec~FWHM.  
We shifted the images into alignment on a fixed star, then computed the median of the set of images.  This median image was then subtracted from the individual images to provide first order removal of the background field. No attempt was made to correct for seeing and tracking differences between the images.  The median-subtracted images were then blinked visually in search of non-sidereal objects.  We found no evidence for R3 in the data, although several faint asteroids (with angular motions distinctly different from those predicted for R3) were evident.  About 5\% of the field was   adversely affected by bright stars, their diffraction patterns and internally-scattered light, leading to a significant local reduction in sensitivity.  In the remainder of the field, the limiting magnitude of the search data was estimated by digitally adding, and then searching for, implanted images of known brightness.  In this way, we estimate a 50\% efficiency for detection in the red 600 s images of 2015 February 17 as $R = 24.4$.  Given the observing geometry (Table \ref{geometry}) and the  color V-R = 0.38$\pm$0.03 (Jewitt et al.~2014), this corresponds to  a limit to the  absolute magnitude $H_V >$ 19.4.    The corresponding cross-section (Equation \ref{area}) is $C_e <$ 0.52 km$^2$, while the effective radius limit is $r_e <$ 0.4 km.    By comparison with Table (\ref{photometry}), we see that the Keck observations would not have been capable of detecting any of the components, all of which have  absolute magnitudes fainter than the Keck limit.  

The diffuse dust envelope must be treated differently.  When extrapolated to the date of the Keck observation, the fit to the cross-section of the diffuse dust shown in Figure (\ref{modelfits}) indicates $C_e \sim$ 14 km$^2$, measured within a circular aperture 8000 km in projected radius.  At $\Delta$ = 2.573 AU, the latter corresponds to a circle 4.3\arcsec~in radius.  By Equations (\ref{absolute}) and (\ref{area}), the apparent magnitude of this dust envelope should then be $V \sim$ 21.2, which would be readily visible if concentrated into the $\sim$0.8\arcsec~seeing disk of the images.  Conversely, spread uniformly over the aperture, the expected dust surface brightness is 25.6 magnitudes (arcsec)$^{-2}$, roughly 5 magnitudes fainter than the background night sky.  This still should be detectable, although much more challenging given the complicated scattered light field in the Keck data.  On this basis, we suspect (but cannot prove) that the late-stage non-detections of R3 result from an erroneous ephemeris rather than from the intrinsic faintness of the dust envelope.  

We also searched for fragments and dust from R3  in images obtained with the FORS2 instrument mounted on the Very Large Telescope (VLT) of the European Southern Observatory (ESO) on Cerro Paranal in Chile. We obtained 30 images of 530\,s exposure time each in R band on 2015 January 18. The covered field of sky is 7.2\arcmin\ $\times$ 7.2\arcmin, and the JPL Horizons ephemeris position is near the upper eastern edge of the field. We obtained a stellar composite by averaging all images in the sidereal reference frame rejecting saturated pixels and the five brightest  and the faintest pixel at each position to exclude cosmic rays and moving objects. The stellar composite was subtracted from each individual exposure and the resulting star-subtracted images were then averaged again in the sidereal frame to detect moving objects in the field of view. We found several moving objects, of which none moved at the rate predicted for R3. The faintest detected moving object was of magnitude $R_{lim}$ = 24.2 $\pm$ 0.1, using the instrumental zero point and nightly extinction coefficient from the ESO database.
At the heliocentric and geocentric distance of R3, the magnitude $R_{lim}$ would have corresponded to an absolute magnitude of $H_V$ = 22.3 $\pm$ 0.1, or an equivalent sphere radius of 110\,m for an albedo of 0.05 by Equation (\ref{area}). We conclude that the field covered by the VLT observations did not contain any R3 fragment larger than 220\,m in diameter.

The star-subtracted images were also averaged in the co-moving frame of R3 to search for a debris trail, using both a sigma-clipping and a minimum-maximum rejection algorithm. We did not find any obvious sign of a debris trail and derive a 1-sigma-per-pixel upper limit surface brightness corresponding to 28.4\,magnitude arcsec$^{-2}$. An extended source of this brightness would be readily visible owing to averaging of the light over many pixels.  For comparison, the debris trails of fragment ejecting asteroids P/2010 A2 (Jewitt et al., 2013b) and 331P/Gibbs (Drahus et al., 2015) had a surface brightness of $\sim$26\,magnitude arcsec$^{-2}$, about 2\,magnitude arcsec$^{-2}$ brighter than our limit.  This is a strong indication that either no comparable quantity of $>$mm-sized debris was  ejected from R3 (in contradiction to our results above), or that the debris trail was not in the field of view of our observations. This adds further support to our conclusion that the non-recovery of R3 after solar conjunction was due to a larger-than-expected ephemeris uncertainty.

\section{DISCUSSION}
The key properties of R3 inferred from the observations described above are as follows.

\begin{enumerate}
\item R3 consists of multiple components separating with a characteristic sky-plane velocity dispersion $\Delta v \sim$ 0.3 m s$^{-1}$.  
\item Measured component break-up times (estimated from fragment motions)  are staggered over $\sim$6 months, indicating that the disintegration is progressive not impulsive.  
\item The individual components of R3 are dust-dominated, so that only upper limits to the sizes of their solid nuclei can be obtained.  These limits are of order $\sim$ 0.1 to 0.2 km (geometric albedo 0.05 assumed).  
\item The best estimate of the upper limit of the radius of the precursor body which broke-up to produce R3 is $\lesssim$ 0.4 km.
\item The system is enveloped in an envelope of dust and debris, with a peak cross-section $\sim$30 km$^2$ that decays on timescales of $\sim$1 month and which is ejected with an initial velocity on the same order as $\Delta v$.  

\item No trace of the object is found in data taken 1 year after the first observations.

\end{enumerate}

What is the cause of the break-up of R3?  Tidal stresses on comets are capable of breaking their nuclei (e.g.~Boehnhardt 2004) but only when inside the Roche lobes of the Sun or a planet.  R3 is dynamically isolated from the Sun and planets and so we dismiss the possibility that it fragmented because of tidal stresses.   

Could R3 have been dispersed by an impact?  That this is possible is shown by existing examples of asteroid - asteroid collision, notably the well-documented impact of a $\sim$35 m scale projectile into the 113 km diameter asteroid (596) Scheila in late 2010 (Bodewits et al.~2011, Jewitt et al.~2011, Ishiguro et al.~2011).  However, the collisional lifetime of a $\sim$400 m radius main-belt asteroid is of order 1.5$\times$10$^8$ yr (Bottke et al.~2005), which is $\sim$10$^2$ times longer than the timescale for spin-up of such a small body caused by radiation forces (the so-called YORP effect, discussed later).  In this sense, impact is less likely than disruption through rotational spin-up.  In addition, the protracted nature of the break-up of R3 argues against an origin by collisional disruption, which we expect to be impulsive in nature rather than spread over a period of many months.   Therefore, while we cannot rule it out, we do not suspect impact disruption as the most likely cause of the break-up of R3.   

The conduction timescale for a spherical body of radius $r_n$ is $\tau_c \sim r_n^2/\kappa$, where $\kappa$ is the thermal diffusivity.  For a porous dielectric solid, we take $\kappa \sim 10^{-7}$ m$^2$ s$^{-1}$ and, substituting $r_n$ = 400 m then gives $\tau_c$ = 1.6$\times$10$^{12}$ s (about 4$\times$10$^4$ years).  Within a small multiple of $\tau_c$, we can safely assume that heat deposited on the surface from the Sun will have conducted all the way to the center, raising the temperature there to approximately equal the local isothermal blackbody temperature, $T_{BB} = 278 r_H^{1/2}$.  At $r_H$ =3 AU, this core temperature is $T_{BB} \sim$160 K.  We calculated the pressure, $P$, produced by the steady-state sublimation of water ice at this temperature, finding $P$ = 4$\times$10$^{-5}$ N m$^{-2}$.  For comparison, the hydrostatic pressure in the core, neglecting rotation, is $P_c \sim 4\pi G \rho^2 r_n^2/3$.  Substituting $\rho$ = 1000 kg m$^{-3}$ and $r_n$ = 400 m, we find $P_c \sim$ 45 N m$^{-2}$.  With $P \ll P_c$, we conclude that gas pressure  from the sublimation of water ice is unable to disrupt R3 against its own gravity.  Super-volatile ices (e.g.~carbon dioxide, carbon monoxide, nitrogen) could generate higher sublimation pressures, if they exist in R3.  For example, the equilibrium  gas pressure produced by carbon monoxide (CO) sublimation at 3 AU is about three orders of magnitude larger at $P_{CO}$ = 2$\times$10$^{-2}$ N m$^{-2}$.  However, this is still $P_{CO} \ll P_c$, meaning that gas pressure is incapable of disrupting R3.  Moreover, while  super-volatile molecules are indeed trapped in comets, their presence reflects  the very low internal temperatures in  bodies  arriving from the frigid Kuiper belt and Oort cloud (30 to 40 K and $\sim$10 K, respectively).  For example, amorphous water ice is the likely carrier of super-volatiles in comets and, while the ice crystallization time exceeds the age of the solar system at 40 K, it is  only $\sim$ 10$^{-6}$ s at 160 K (Jenniskens et al.~1998).  As a result, neither amorphous ice nor super-volatiles can be retained at  asteroid temperatures (Prialnik and Rosenberg 2009). For all these reasons it is difficult to see how gas pressure alone could disrupt R3.

By elimination, then, rotational instability offers the most plausible cause of break-up in R3, as suggested earlier by Jewitt et al.~(2014) and modeled by Hirabayashi et al.~(2014).  Rotational breakup of asteroids is well-established where, for all but the smallest, there exists a rotation barrier near periods of $\sim$2.2 hour (e.g.~Warner et al.~2009, Chang et al.~2015).  Asteroids rotating faster than this critical value are presumed to have been destroyed when centrifugal forces overcame the gravitational and cohesive forces binding them together.  Asteroids are thought to have very small cohesive strengths as a consequence of pre-fracturing by impacts having too little energy to cause dispersion of the fragments, forming (``rubble piles'').  Block-rich, 500 m scale asteroid (25143) Itokawa provides our closest view of such a rubble-pile object (Fujiwara et al.~2006), and may be a useful analog to the precursor of R3.  The early-time distribution of the A, B and C components of R3 (Figures \ref{october02}, \ref{october28} and \ref{whole}) is along a line that is parallel to neither the projected orbit nor the antisolar direction, perhaps marking the projected rotational equator of the parent.


The rotational break-up of a  body should release fragments that separate with initial speeds dependent on the bulk density, shape and cohesive strength (e.g.~Van wal and Scheeres 2016).  All else being equal, the greater the cohesion the larger the angular frequency needed to induce breakup and the larger the break-up separation speeds.  The tiny fragment velocity dispersion in R3, $\Delta v =$ 0.33$\pm$0.03 m s$^{-1}$, immediately implies a small cohesive strength given, to order of magnitude,  by Equation (5) of Jewitt et al.~(2015a)

\begin{equation}
S \sim \rho \left(\frac{r_s}{r_p}\right) (\Delta v)^2.
\label{strength}
\end{equation}

\noindent Here, both components are assumed to be  of the same bulk density, $\rho$, and geometric parameters associated with the body shape are ignored.  Substituting $\rho$ = 10$^3$ kg m$^{-3}$, $(r_s/r_p) \sim$ 1/2 to 1 and $\Delta v$ = 0.33 m s$^{-1}$ we obtain $S \sim$ 50 to 100 N m$^{-2}$.  A more sophisticated model (but still relying on observationally uncertain parameters including the size, density and shape of the components) gives an overlapping range 40 $\le S \le$ 210 N m$^{-2}$ (Hirabayashi et al.~2014).    The important conclusion is that the cohesive strengths implied by the measured $\Delta v$ are far smaller than the values representative of competent rocks (10$^7$ to 10$^8$ N m$^{-2}$) and comparable to the cohesion resulting from van der Waals forces in fine powders, as expected in a heavily fractured  body (Sanchez and Scheeres 2014). 

If R3 is a purely rocky asteroid, the torque needed to provide rapid rotation must be contributed by solar radiation through the YORP effect.  The YORP e-folding timescale, $\tau_Y$, is a function of the size, shape, spin vector, thermal properties and detailed structure of the surface materials.  Simplistic treatments show that $\tau_Y$ scales as  

\begin{equation}
\tau_Y \sim \psi r_n^2  r_H^2,
\label{yorp}
\end{equation}

\noindent where $\psi$ is a constant and $r_n$ and $r_H$ are the asteroid radius (in meters) and heliocentric distance (in AU), respectively.  The constant of proportionality  in this relation cannot be accurately calculated a-priori for any given asteroid (because it is a function of many unknowns), but can be estimated retroactively from measurements of the YORP effect in well-characterized asteroids  (e.g.~Lowry et al.~2014).   In this way we find $\psi$ = 1.3$\times$10$^7$ s m$^{-2}$ AU$^{-2}$ (Equation (3) of Jewitt et al.~2015a).     The precursor to R3, with  $r_n =$ 400 m and orbiting at $r_H \sim$ 3.2 AU (the effect of R3's modest eccentricity is negligible, for our present purposes) has $\tau_Y \sim 2\times 10^{13}$ s  (0.7 Myr), by Equation (\ref{yorp}).  If R3 is a nearly strengthless aggregate body then, within a small multiple of $\tau_Y$, it should be spun-up to the point of disruption.  The YORP timescale is two orders of magnitude smaller than the likely collisional age ($\sim$150 Myr) of a 0.4 km main-belt asteroid, showing that YORP torque spin-up is entirely plausible for this object.

If R3 contains ice, as do some other active asteroids (specifically 133P/Elst-Pizarro; Hsieh and Jewitt 2006, 238P/Read; Hsieh et al.~2011, 313P/Gibbs; Jewitt et al.~2015b and 324P/La Sagra; Hsieh and Sheppard 2015) then sublimation torques instead could drive the spin-up.  Torques due to non-central outgassing forces have long been recognized as capable of causing the spin-up and even break-up of cometary nuclei (e.g.~Jewitt 1992, 1997).   While no evidence for outgassing has been found in R3 (the limit to the water ice sublimation rate is set spectroscopically at $dM/dt \le 1$ kg s$^{-1}$; Jewitt et al.~2014), it is reasonable to consider the possible role of sublimation torques in the break-up. 

\begin{figure}
\epsscale{0.95}
\begin{center}
\includegraphics[width=0.95\textwidth, angle =0 ]{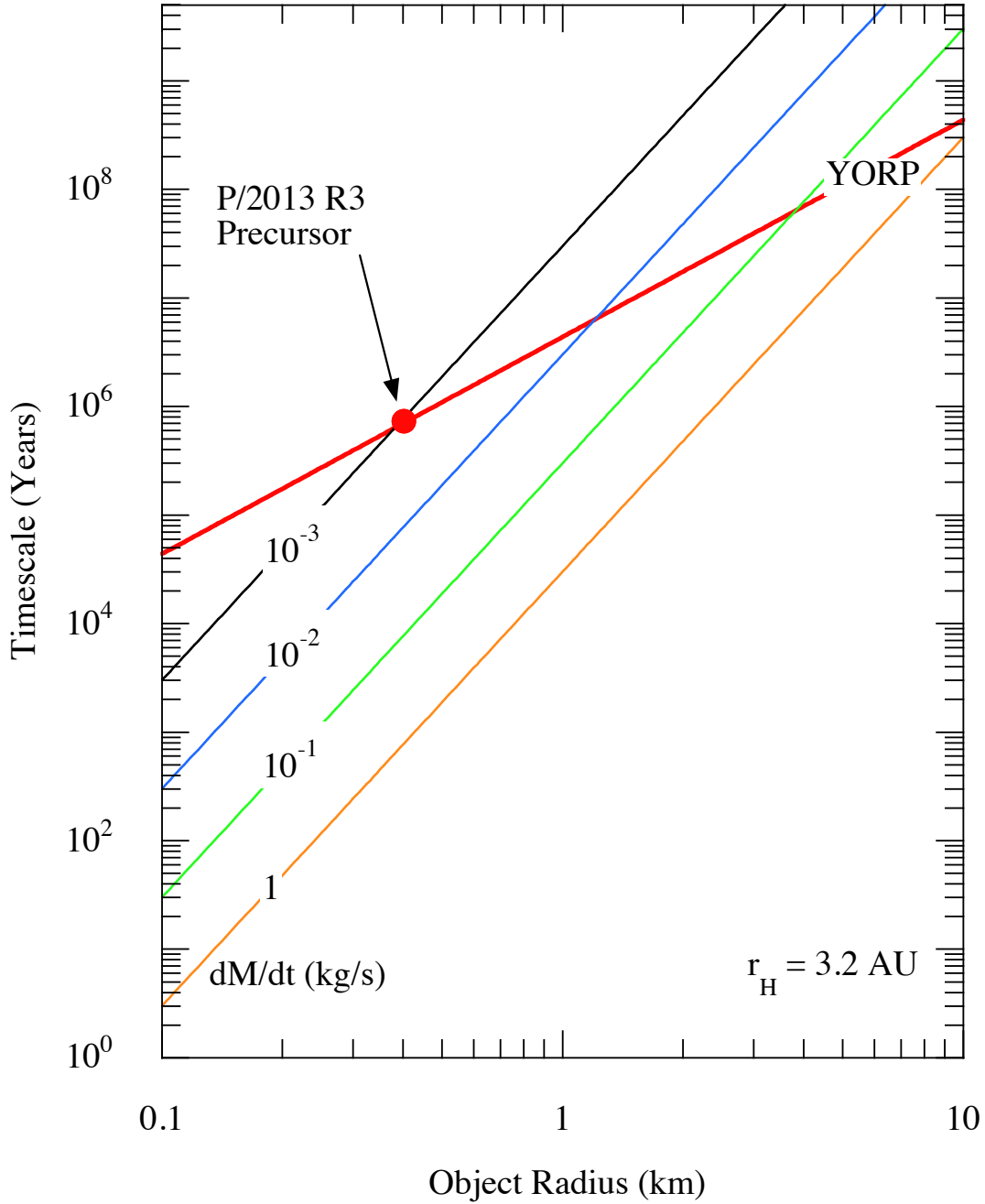}

\caption{The e-folding timescales for spin-up by the YORP effect (thick red line) and sublimation torques (steeper colored lines, each labeled with the mass loss rate in kg s$^{-1}$, from Equations (\ref{yorp}) and (\ref{spin1}). All calculations apply at $r_H$ = 3.2 AU. \label{timescales}
} 
\end{center} 
\end{figure}

The timescale for spin-up caused by outgassing at rate $dM/dt$ kg s$^{-1}$ is (Jewitt et al.~2016)

\begin{equation}
\tau_s \sim \frac{\omega \rho r_n^4}{V_{th} k_T (dM/dt)}, 
\label{spin1}
\end{equation}

\noindent where $\omega$ is the initial angular frequency of the rotation, $\rho$ is the bulk density, $V_{th}$ is the speed of sublimated molecules, and $0 \le k_T \le 1$ is a dimensionless number representing the moment arm of the outgassing torque.    To estimate $\tau_s$, we assume an initial rotation period of 5 hours (i.e.~$\omega = 3.5\times 10^{-4}$ s$^{-1}$), $\rho$ = 10$^3$ kg m$^{-3}$, $V_{th}$ = 400 m s$^{-1}$ (the speed of sound in H$_2$O gas at the 155 K blackbody temperature appropriate to 3 AU) and use $k_T = 10^{-3}$, as suggested by measurements of the spin-up of comets.   These results are plotted in Figure (\ref{timescales}).  

Comparing $\tau_Y$ with $\tau_s$, we find that outgassing torques  exceed the YORP torque provided the mass loss rate exceeds a critical value given by 

\begin{equation}
\frac{dM}{dt} > \frac{\omega \rho r_n^2}{\psi r_H^2 V_{th} k_T}. 
\end{equation}

\noindent  With the substitutions used above, we obtain 

\begin{equation}
\frac{dM}{dt} > 7\times10^{-9} r_n^2
\end{equation}

\noindent such that the R3 precursor with $r_n$ = 400 m corresponds to $dM/dt > 10^{-3}$ kg s$^{-1}$.  This tiny mass loss rate is three orders of magnitude smaller than the empirical limit to the mass loss (1 kg s$^{-1}$) based on observations (Jewitt et al.~2014).   It could be generated by equilibrium sublimation of exposed ice from as little as a few tens of m$^2$ of surface (5$\times$10$^{-6}$ of the surface of a 400 m radius sphere).  We conclude that unobservably small sublimation mass loss rates, if sustained, can rival or surpass the YORP torque.  

Whatever the source of the torque,  rotational instability of a low strength body does appear to match the gross characteristics of R3.
Simulations show that the process of break-up of an aggregate body having gravity and strength has many surprising complexities.  For example, the shape of an aggregate (and the resulting torque vector) can adjust as the spin increases, even before material is lost (Cotto-Figueroa et al.~2015). Once lifted from the surface,  slowly-launched fragments can fall back, become trapped in orbit or escape depending on complex gravitational and collisional interactions that lead to the inherent unpredictability of dynamical chaos  (Boldrin et al.~2016, Sanchez and Scheeres 2016). A hint of this is especially clear in the models of Boldrin et al.~(2016), in which the effects of non-zero obliquities of the components in a disrupting, non-planar binary are explored.  There, dynamical instabilities persist (for high mass ratio systems) for 100s and even 1000s of days, and include secondary fission in about 1/6th of the cases considered.   The observed  secondary fragmentation in R3 (e.g.~of component A1 into A2, A4, A5 and A6) and even tertiary fragmentation (e.g.~A2 into A3), with intervals from $\sim$10 to $\sim$100 days, may reflect the continued instability of weakly bound clods ejected in excited rotational states by the initial break-up event.  In this picture, the separation velocities in R3 reflect gravitational scattering more than material strength, and  Equation (\ref{strength}) can give only an upper limit to the cohesion of the parent body.

\subsection{Mass Flux into the Zodiacal Cloud}
To estimate the rate of supply of debris to the Zodiacal Cloud complex from R3-like disruptions in the asteroid belt we need to know the rate at which these events occur.  This is difficult to estimate from the available data because the detection parameters (areal coverage, cadence, depth) of a majority of the current surveys that are finding active asteroids remain poorly characterized in the published literature (the main exception is Panstarrs; Denneau et al.~2015). We assume that surveys capable of finding R3 have efficiency, $\varepsilon$,  that such surveys have been in operation for a time, $t$, and that only one disruption (R3) of mass $M$ = 2.7$\times$10$^{11}$ kg has so far been detected.  We write  $dM/dt \sim M/(\varepsilon t)$.   To set a lower limit to the mass input rate, we set $\varepsilon$ = 1 and $t$ = 10 years, to find  $dM/dt \ge$ 1000 kg s$^{-1}$. This is a lower limit because, in reality, no survey can operate at 100\% efficiency (the day-night cycle, lunar interference and weather losses mean that even $\varepsilon$ = 50\% is rarely possible) and the most powerful surveys (Catalina Sky Survey, Panstarrs) have been operating at full sensitivity for a  time substantially shorter than 10 years.

For comparison, the largest published estimate of the steady-state rate of supply needed to maintain the Zodiacal Cloud is  $dM_{ZC}/dt = 2.5\times10^4$ kg s$^{-1}$ (Table 1 of Nesvorny et al.~2011), giving the fractional contribution from asteroids as $(dM/dt)/(dM_{ZC}/dt) \ge$ 0.04.  Nesvorny's preferred estimate,  $dM_{ZC}/dt = 5\times10^3$ kg s$^{-1}$ (David Nesvorny, personal communication), gives $(dM/dt)/(dM_{ZC}/dt) \ge$ 0.2.  Using other methods, Yang and Ishiguro (2015) modeled the color of the Zodiacal Cloud dust to infer that  $\lesssim$10\% of the particles are from asteroids.  Separately, Carrillo-S{\'a}nchez et al.~(2016) used an atmospheric ablation model and the abundance of cosmic spherules to conclude that $\sim$8\% of the interplanetary dust striking Earth has an asteroidal source.  Our estimate is consistent with these independent values.  We concur with Nesvorny et al.~(2011) that asteroid dust is a  measurable, but probably not dominant, contributor to the Zodiacal Cloud.  Future sky surveys, with better-defined efficiencies, $\varepsilon$, and durations, $t$, will enable us to determine the asteroidal contribution with more confidence.

\clearpage

\section{SUMMARY}

We present an analysis of the full suite of available observations of disintegrating asteroid P/2013 R3 from ground-based and space-based telescopes.  

\begin{enumerate}

\item The data are consistent with P/2013 R3 being an aggregate body (initial radius $\sim$400 m), driven to rotational instability in 2013 August 13$\pm$24.  Torques from radiation and, if ice is present, from anisotropic sublimation  are easily capable of driving a 400 m body to break-up in a time short compared to the collisional lifetime.

\item The breakup was accompanied by the release of an extensive debris cloud having a peak cross-section $\sim$30 km$^2$, a minimum particle size measured in millimeters, and an inferred mass $\sim$2$\times$10$^{10}$ kg (density 10$^3$ kg m$^{-3}$ assumed).  

\item First generation fragments from the initial breakup show further fragmention accompanied by the release of dust in the 5 month interval from 2013 August to 2014 January. The  mean pair-wise sky-plane velocity dispersion between fragments, $\Delta v$ = 0.33$\pm$0.03 m s$^{-1}$,  indicates an effective cohesive strength $\sim$50 to 100 N m$^{-2}$.   

\item P/2013 R3 style rotational disruptions supply a measurable but probably not dominant fraction of the dust needed to maintain the Zodiacal Cloud in steady-state.  We estimate a rate $dM/dt \gtrsim$10$^3$ kg s$^{-1}$, corresponding to $\gtrsim$4\% of the total Zodiacal Cloud mass flux.

\item Non-detections of P/2013 R3 in 2015 likely reflect an inaccurate ephemeris.

%
%
%
%
%

\end{enumerate}

\acknowledgments
We thank Scott Sheppard and Man-To Hui for comments and Michal Drahus for a detailed review. Based in part on observations made with the NASA/ESA \emph{Hubble Space Telescope,} with data obtained at the Space Telescope Science Institute (STSCI).  Support for programs 13612 and 13865 was provided by NASA through a grant from STSCI, operated by AURA, Inc., under contract NAS 5-26555.  We thank Linda Dressel, Alison Vick and other members of the STScI ground system team for their expert help.   Some of the data presented herein were obtained at the W.M. Keck Observatory, operated as a scientific partnership among Caltech, the University of California and NASA. The Observatory was made possible by the generous financial support of the W. M. Keck Foundation. Based in part on observations made with ESO Telescopes at the La Silla Paranal Observatory under program ID 094.C-0742(A).  DJ appreciates support of a grant from NASA's Solar System Observations program.

\clearpage

\begin{deluxetable}{llcccrrcr}
\tablecaption{Observing Geometry 
\label{geometry}}
\tablewidth{0pt}
\tablehead{ \colhead{Tel} & \colhead{UT Date and Time} & DOY\tablenotemark{a}   & \colhead{$r_H$\tablenotemark{b}}  & \colhead{$\Delta$\tablenotemark{c}} & \colhead{$\alpha$\tablenotemark{d}}   & \colhead{$\theta_{\odot}$\tablenotemark{e}} &   \colhead{$\theta_{-v}$\tablenotemark{f}}  & \colhead{$\delta_{\oplus}$\tablenotemark{g}}   }
\startdata
Keck & 2013 Oct 01  07:45 - 08:20 &   274  &  2.230 & 1.230  & 1.7 & 235.7 & 246.2  & -0.31\\
Keck & 2013 Oct 02  07:17 - 09:50 &   275   &  2.231 & 1.231  & 1.2 & 230.5 & 246.2  & -0.33\\
Magellan & 2013 Oct 28 00:31 - 00:40 & 301 & 2.260 & 1.331 & 11.6 & 68.5 & 245.8 & -0.53 \\
Magellan & 2013 Oct 29 00:29 - 00:35 & 302 & 2.262 & 1.338 & 12.1 & 68.4 & 245.8 & -0.54 \\
HST & 2013 Oct 29  06:36 - 08:17 &   302 &     2.262 & 1.338  & 12.1 & 68.4 & 245.8  & -0.54\\
HST & 2013 Nov 15   06:39 - 07:20  &   319 &    2.287 & 1.489  & 18.2 & 67.6 & 245.7  & -0.56\\
HST & 2013 Dec 13   07:25 - 08:05   & 347   &    2.335 & 1.827 & 23.5 & 67.2 & 245.9 & -0.48 \\
HST & 2014 Jan 14   09:24 - 10:04  & 379 &  2.402 & 2.281 & 24.1 & 67.6 & 246.8 & -0.29 \\
HST & 2014 Feb 13 09:52 - 10:33 & 409 &  2.472 & 2.712 & 21.3 & 69.0 & 248.7 & -0.10 \\
HST & 2014 Sep 29 01:02 - 01:40  &    637 & 3.090 & 3.331 &   17.4 & 281.9  & 282.3   & -0.10\\
HST & 2014 Oct 28 00:54 - 01:31 &    666 & 3.165&  2.997 &   18.3 & 283.6  & 284.5   & -0.24 \\
HST & 2014 Dec 09 18:20 - 18:57 &    708 &  3.272&  2.526 &   12.8 &  283.3  & 285.0   & -0.35\\
HST & 2015 Jan 17 12:02 - 17:22 & 747 & 3.362 & 2.378 &    0.6 & 255.9  & 282.5   & -0.27\\
VLT  & 2015 Jan 18 02:45 - 07:43          & 748 & 3.363  & 2.379 & 0.4   & 238.8  & 282.5    & -0.27 \\
Keck & 2015 Feb 17  06:18 - 07:20 & 778 & 3.429 & 2.573 & 9.5 & 100.8 & 280.3 & -0.09\\
HST & 2015 Mar 04 04:45 - 05:22 &    766 & 3.459 &  2.750 &   12.9 &   99.7  & 279.7   & -0.00 \\
HST & 2015 Apr 07 12:56 - 13:34 &    827 & 3.527 &  3.283 &   16.4 &   99.6  & 280.2    & 0.15 \\
HST & 2015 May 26 04:50 - 05:27 &    876 & 3.612 & 4.052 &  13.7 & 102.7  & 283.6    & 0.21 \\
Keck & 2015 Dec 08  13:21 - 14:30 & 1072 &  3.837 & 3.842 & 14.7 & 293.3 & 294.3 & -0.23 \\

\enddata


\tablenotetext{a}{Day of Year, UT 2013 January 01 = 1}
\tablenotetext{b}{Heliocentric distance, in AU}
\tablenotetext{c}{Geocentric distance, in AU}
\tablenotetext{d}{Phase angle, in degrees}
\tablenotetext{e}{Position angle of the projected anti-Solar direction, in degrees}
\tablenotetext{f}{Position angle of the projected negative heliocentric velocity vector, in degrees}
\tablenotetext{g}{Angle of Earth above the orbital plane, in degrees}

\end{deluxetable}

\clearpage

\begin{deluxetable}{clrrllccc}
\tablecaption{Pair-Wise Separation Speeds and Dates
\label{pairs}}
\tablewidth{0pt}
\tablehead{ \colhead{Parent} & \colhead{Child} &\colhead{$\Delta v$ (m s$^{-1})$}&  \colhead{ DOY\tablenotemark{a}} & \colhead{UT Separation Date}  }
\startdata
A1 & A2  & 0.23$\pm$0.05 		& 290$\pm$10 & 2013 October 17 \\
A1 & A4 & 0.33$\pm$0.05 		& 298$\pm$10 & 2013 November 03 \\
A1 & A5 & 0.33$\pm$0.05 		& 309$\pm$10 & 2013 November 5 \\
A1 & A6 & 0.31$\pm$0.05 		& 294$\pm$10 & 2013 October 20 \\
A2 & A7 & 0.46$\pm$0.05 		& 328$\pm$10 & 2013 November 24 \\
A1? & D   & $<$6 (3$\sigma$)             & $<$211           & $<$ 2013 July 30 \\
B1& B2  	& 0.32$\pm$0.02  		&	319$\pm$10 	&   2013 November 15 \\
B1& B3 & $\ge$0.28 			& $>$379 	& $>$2014 January 14 \\

\enddata


\tablenotetext{a}{Estimated date of separation of the components expressed as Day of Year (2013)}

\end{deluxetable}

\clearpage

\begin{deluxetable}{llcccccccc}
\tablecaption{Dust Tail Position Angles
\label{positionangle}}
\tablewidth{0pt}
\tablehead{ \colhead{UT Date} & \colhead{Telescope}&  \colhead{Feature\tablenotemark{a}} & \colhead{East Tail} & West Tail       }
\startdata
2013 Oct 01   	& Keck & Diffuse &	--	&   243.6$\pm$0.4 \\
2013 Oct 02   	&Keck & Diffuse&	--	&   244.9$\pm$0.3 \\
2013 Oct 28,29   & Magellan	& Diffuse &  69.4$\pm$1.4 & 240.2$\pm$0.2 \\
2013 Oct 29  & HST  & Diffuse &   72.5$\pm$1.2 & 240.1$\pm$0.3 \\

2013 Nov 15   & HST 	&  A & 69.8$\pm$0.8 & -- \\
2013 Nov 15   & HST 	&  B & 68.2$\pm$0.5 & -- \\
2013 Dec 13   & HST  &  A & 65.4$\pm$1.3 & --\\
2013 Dec 13   & HST  &  B & 70.4$\pm$1.5 & --\\
2014 Jan 14    & HST &  Diffuse & 66.9$\pm$0.6 & -- \\
2014 Feb 13  & HST&  Diffuse & 72.6$\pm$0.6 & --\\

\enddata


\tablenotetext{a}{``Diffuse'' refers to large dust structures enveloping the multiple nucleus system.  Dust structures associated with specific components are labeled A and B, as appropriate,}

\end{deluxetable}

\clearpage

\begin{deluxetable}{lcccccc}
\tablecaption{Hubble Telescope Photometry
\label{photometry}}
\tablewidth{0pt}
\tablehead{
\colhead{Name}    & \colhead{Quantity\tablenotemark{a}}   & \colhead{Oct 29}     & \colhead{Nov 15} & \colhead{Dec 13} & \colhead{Jan 14} & \colhead{Feb 13}\\
}

\startdata

A1 & V & 23.38$\pm$0.02 	& 24.31$\pm$0.03  & 24.91$\pm$0.03 & 25.75$\pm$0.10 & 26.01$\pm$0.05\\
     & H$_V$ &  20.24$\pm$0.02	& 20.71$\pm$0.03 		& 20.65$\pm$0.03 		& 20.93$\pm$0.10 & 20.84$\pm$0.05\\
    & $C_e$/$r_e$ &  0.24/0.28	& 0.16/0.22	          & 0.16/0.23	& 0.13/0.20 & 0.14/0.21\\
    
A2 & V & 23.95$\pm$0.03  	& 24.40$\pm$0.03 	& 25.05$\pm$0.05 & 25.78$\pm$0.06 & 26.40$\pm$0.17\\
    & H$_V$ &   20.81$\pm$0.03	& 20.80$\pm$0.03 		& 20.79$\pm$0.05 		& 20.96$\pm$0.06 & 21.23$\pm$0.17\\
    & $C_e$/$r_e$ &  0.14/0.21			& 0.14/0.21  		&  0.14/0.21 		& 0.12/0.20 & 0.10/0.18 \\
    
B & V & 22.20$\pm$0.02 		& 23.67$\pm$0.02 & -- & -- & -- \\
    & H$_V$ & 19.06$\pm$0.02		& 20.07$\pm$0.02  &  -- & -- & -- \\
    & $C_e$/$r_e$ &  0.71/0.48			& 0.28/0.30  &  -- & &-- \\
    
B1 & V & -- 				& -- &  24.89$\pm$0.02 & 24.34$\pm$0.14 & 24.76$\pm$0.14\\
     & H$_V$ & -- 				& -- &  20.63$\pm$0.02 & 19.52$\pm$0.14 & 19.59$\pm$0.14 \\
    & $C_e$/$r_e$ &  --		& -- &   0.17/0.23  & 0.47/0.39 &  0.44/0.37\\
    
B2 & V & -- 				& -- & 25.34$\pm$0.03 & 25.96$\pm$0.06 & 26.31$\pm$0.11\\
     & H$_V$ &  -- 			& -- & 21.08$\pm$0.03  & 21.14$\pm$0.06 & 21.14$\pm$0.11\\
    & $C_e$/$r_e$ &  --		& -- & 0.11/0.19 & 0.10/0.18 & 0.10/0.18 \\
    
C1 & V & 26.73$\pm$0.14  	& -- & --  & -- & -- \\
    & H$_V$ &  23.59$\pm$0.14 	& --& -- & -- & -- \\
    & $C_e$/$r_e$ &  	0.01/0.06		& -- &  -- & -- & -- \\

C2 & V & 25.05$\pm$0.02  	& 25.47$\pm$0.06 	& 26.55$\pm$0.03 & 27.27$\pm$0.18 & 28.00$\pm$0.30\\
     & H$_V$ &  21.91$\pm$0.02	& 21.87$\pm$0.06 		& 22.29$\pm$0.03 & 22.45$\pm$0.18 & 22.83$\pm$0.30 \\
    & $C_e$/$r_e$ &  	0.05/0.13			& 0.05/0.13 		&  0.04/0.11 & 0.03/0.10 & 0.02/0.08 \\
    
Diffuse & V & 18.03 & 18.60 & 19.47 & 20.14 & 20.59  \\
           & H$_V$ & 14.89 & 15.00 & 15.21 & 15.32 & 15.42 \\
           & $C_e$/$r_e$ & 33.2/3.3 & 30.0/3.1 & 24.7/2.8 & 22.3/2.7 & 20.4/2.5 \\

\enddata


\tablenotetext{a}{The apparent V magnitude measured within a circular aperture of linear radius $195$ km at the comet, except for the diffuse magnitude, extracted from a region 8000 km in radius. $H_v$ is the corresponding absolute magnitude from Equation (\ref{absolute}).  $C_{e}$ is the effective area in km$^2$ (Equation \ref{area}).  $r_e = (C_E/\pi)^{1/2}$ is the radius of the equal area circle, in km.}

\end{deluxetable}

\clearpage

\clearpage

\clearpage

\clearpage

\clearpage

\clearpage

\clearpage

\clearpage

\clearpage

\clearpage

\clearpage

\clearpage

\clearpage

\clearpage

\end{document}